\documentclass[fontsize=10.5pt,headings=small,headinclude=true,footinclude=true,twocolumn, % twocolumn auskommentieren
 headings=optiontohead,toc=listof,DIV=25 % calc
,listof=leveldown, headsepline, footsepline ]{scrartcl}
\usepackage[utf8]{inputenc}
\usepackage[T1]{fontenc}
\usepackage[ngerman, english]{babel}

\usepackage{tocloft}

\newcommand{\listfootnotesname}{List of Todos}% 'List of Footnotes' title 
\newlistof{footnotes}{fnt}{\listfootnotesname}% New 'List of...' for footnotes 

\newcommand{\dptitle}{Justice as a Social Bargain and Optimization Problem} 

\newcommand{\dpautoren}{Andreas Siemoneit \href{https://orcid.org/0000-0003-4870-8481}{\includegraphics[width=1em]{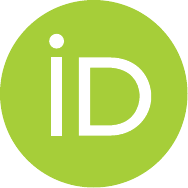}}}

\newcommand{\dpaffiliation}{Carl von Ossietzky Universität Oldenburg, Germany. \par ZOE Institute for future-fit economies, Bonn, Germany.}

\usepackage{amsmath}
\usepackage{amssymb,amsfonts,textcomp}
\usepackage{color}
\usepackage{dpfloat}
\usepackage{array}
\usepackage{txfonts}
\usepackage{booktabs}
\usepackage{paralist}
\usepackage{hhline}
\usepackage{graphicx}
\usepackage{hanging}
\usepackage{bm} \usepackage{bbold}
\usepackage[figuresleft]{rotating}
\usepackage{acronym}
\usepackage{microtype}
\usepackage[hang]{footmisc}
\usepackage{changepage} %for moving tables
\usepackage{appendix}
\usepackage{csquotes}

\usepackage[backend=biber, natbib=true, style=authoryear-comp, sorting=ynt , url=true, isbn=false, doi=true, eprint=false, maxcitenames=2, citetracker=true, uniquename=false]{biblatex}

\DeclareSourcemap{%maps the language to hyphenation
 \maps{
 \map{
 \step[fieldsource=language, fieldset=langid, origfieldval, final]
 \step[fieldset=language, null]
 }
 }
}

\usepackage[inner=1.5cm,outer=1.5cm,top=0.9cm,bottom=1.4cm,includeheadfoot , heightrounded]{geometry} % use for twopage

\usepackage{chngcntr}
\usepackage[hidelinks, breaklinks=true, 
pdfauthor={Andreas Siemoneit}, pdftitle={\dptitle}]{hyperref}

\makeatletter
\newcommand\arraybslash{\let\\\@arraycr}
\makeatother

\makeatletter
\g@addto@macro\UrlBreaks{\do\*\do\~\do\'\do\"\do\a\do\b\do\c\do\d\do\e\do\f\do\g\do\h\do\i\do\j\do\k\do%
\l\do\m\do\n\do\o\do\p\do\q\do\r\do\s\do\t\do\u\do\v\do\w\do\x\do\y\do\z\do\&\do\1\do\2\do\3\do\4\do\5\do\6\do\7\do\8\do\9\do\0\do\.}
\makeatother
\DeclareFieldFormat{url}{\url{#1}}

\usepackage{scrlayer-scrpage}

\clearscrheadfoot
\ihead{\pagemark} 
\ofoot[%Institut für zukunftsfähige Ökonomien $\cdot$ Discussion Paper \dpnumber /\dpyear
]{}
\ifoot[%\href{https://www.zoe-institut.de/dp\dpnumber}{www.zoe-institut.de\,/\,dp\dpnumber}
]{}% \ohead{} %{\pagemark}
\let\OLDthebibliography\thebibliography
\renewcommand\thebibliography[1]{
 \OLDthebibliography{\#1}
 \setlength{\parskip}{0pt}
 \setlength{\itemsep}{0pt plus 0.3ex}
}

\KOMAoption{listof}{leveldown}

\usepackage{ragged2e}
\newcolumntype{L}[1]{>{\raggedright\arraybackslash}p{\#1}} % linksbündig mit Breitenangabe
\newcolumntype{C}[1]{>{\centering\arraybackslash}p{\#1}} % zentriert mit Breitenangabe
\newcolumntype{R}[1]{>{\raggedleft\arraybackslash}p{\#1}} % rechtsbündig mit Breitenangabe

\newcommand{\dpnumber}{7} %%% hier die Nummer des DP einfügen
\newcommand{\dpyear}{2021} %%% hier das Jahr einfügen
\newcommand{\dpmonat}{June} %%% Januar bzw. February, je nach Sprache :)

\ohead{Siemoneit: \dptitle}

\begin{document}

%%% english oder ngerman
\selectlanguage{english}

\begin{titlepage}
\thispagestyle{scrplain}
\enlargethispage{\baselineskip}

% \twocolumn[{%

% \noindent { % \Huge \textcolor{ashgrey}{\textsf{\emph{ZOE Discussion Papers}}} \hfill \includegraphics[width=0.35\textwidth]{ZOE_Logo.png} \par } 

\bigskip

\vspace{0.5em}\noindent\includegraphics[width=\textwidth]{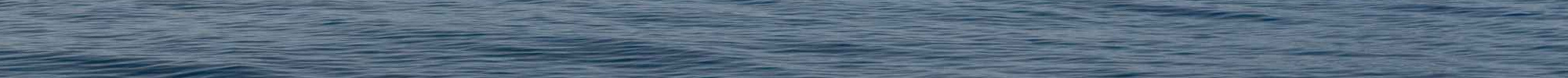} \par   

\vspace{0.7em} 
\begin{center}

 { \Large { \bfseries \sffamily \dptitle \par } \vspace{1em} {\large \dpautoren \par \vspace{1em} \normalsize \dpaffiliation \par } } 
\vspace{1.3em}

 %{ \large Draft of a Journal Article\ $\cdot$\ \dpmonat~\dpyear } 

{ \large \dpmonat~\dpyear } 

\end{center}

\vspace{1.5em}

\begin{addmargin}{0.15\textwidth}

%%% hier abstract einfügen
\textbf{Abstract:} 
The question of ``Justice'' still divides social research and moral philosophy. 
Several Theories of Justice and conceptual approaches compete here, and distributive justice remains a major societal controversy. 
From an evolutionary point of view, fair and just exchange can be nothing but ``equivalent'', and this makes ``strict'' reciprocity (merit, equity) the foundational principle of justice, both theoretically and empirically. 
But besides being just, justice must be effective, efficient, and communicable.
Moral reasoning is a communicative strategy for resolving conflict, enhancing status, and maintaining cooperation, thereby making justice rather a social bargain and an optimization problem.
Social psychology (intuitions, rules of thumb, self-bindings) can inform us when and why the two auxiliary principles \emph{equality} and \emph{need} are more likely to succeed than \emph{merit} would. 
Nevertheless, both equality and need are governed by reciprocal considerations, and self-bindings help to interpret altruism as ``very generalized reciprocity''.
The \emph{Meritocratic Principle} can be implemented, and its controversy avoided, by concentrating on ``non-merit'', i.e., institutionally draining the wellsprings of undeserved incomes (economic rents).
Avoiding or taxing away economic rents is an effective implementation of justice in liberal democracies. 
This would enable market economies to bring economic achievement and income much more in line, thus becoming more just.

\bigskip

%%% Keywords / Schlagwörter

\bigskip \bigskip

%%% Creative Commons Lizenz, falls gewünscht
 \noindent\begin{minipage}[t]{0.84\linewidth}
 \textbf{License:} Creative-Commons \href{http://creativecommons.org/licenses/by-nc-nd/4.0/}{CC-BY-NC-ND 4.0}. \end{minipage} 
 \begin{minipage}[t]{0.15\linewidth}\vspace{-\ht\strutbox}
 \includegraphics[width=\columnwidth]{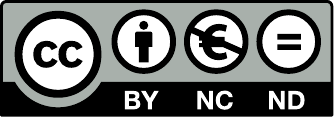}\end{minipage}

\bigskip

\noindent\textbf{Highlights:}

\begin{itemize}
    \item Reciprocity is primary principle of justice, only complemented by need and equality
    \item Justice must be just, but also effective, efficient, and communicable
    \item Need and equality are governed by reciprocal considerations
    \item Self-bindings help to interpret altruism as ``very generalized reciprocity''
    \item Avoiding or taxing away economic rents is an effective implementation of justice
\end{itemize}

\end{addmargin}

%\listoffootnotes

\vfill

{

% \noindent \textbf{Herausgeber\,/\,Publisher:} ZOE. Institut für zukunftsfähige Ökonomien e.\,V., Thomas-Mann-Straße 36, 53111 Bonn. info@zoe-institut.de $\cdot$ www.zoe-institut.de $\cdot$ ISSN 2627-9436. \\ \noindent \foreignlanguage{ngerman}{ZOE Diskussionspapiere stellen Forschungsergebnisse und Thesen für eine sozial-ökologische Wirtschaft und Gesellschaft vor.} \foreignlanguage{english}{ZOE discussion papers present research results and theses for a socio-ecological economy and society.}

}

\clearpage

\begin{addmargin}{0.11\textwidth}

\ 
\vspace{1cm}

\tableofcontents

\end{addmargin}

\end{titlepage}

\pagestyle{scrheadings}

\section{Introduction}
\label{justsec_introduction}

The scientific and philosophical debate about justice has not yet revealed a clear paradigm, with several Theories of Justice competing here \citep{cohen_justice:_1986, sandel_justice:_2009, miller_justice_2017}:
\begin{itemize}
    \item \textbf{Grand Theories of Justice}, usually attributed to some founder \citep[e.g.,][]{sandel_justice:_2009}: Virtue -- Aristotle, Utilitarianism -- Bentham, Autonomy (Deontology) -- Kant.
    \item \textbf{Basic principles} that would (alone or in combination) govern the human quest for justice: virtue, happiness, merit, sufficiency, priority, need, equality, liberty, self-ownership etc. 
    While most theories are devoted to a primary principle of justice, some of them make a case for a plurality of principles (\citealt{walzer_spheres_1983, audi_moral_2002}, see also below on the ``tripartite approach'').
    \item Different \textbf{conceptual approaches}, for example naturalism (tracing justice back to natural phenomena -- utilitarianism, intuitionism), contractarianism (justice being a kind of social contract), libertarianism (an extreme form of individualism), and several others \citep[cf.][]{olsaretti_handbook_2018}.  
\end{itemize}

We find major conceptual contrasts (tensions, dichotomies). Among these are:
\begin{itemize}
    \item \textbf{Substantive vs.\ procedural} justice \citep[e.g.,][]{miller_justice_2017}: Can justice strive for certain desired outcomes, or must justice restrict itself to just procedures, accepting any outcome? 
    More generally, social psychologists are concerned with how procedural questions (non-outcome factors) affect perceptions of justice \citep{lind_study_2020}.
    \item \textbf{Cognitivist vs.\ decisionist} approaches \citep[e.g.,][40ff.]{quante_einfuhrung_2013}: Is justice based on principles to be discovered or (only) subject to contingent agreement?
    A similar contrast is \textbf{rationalist vs.\ empiricist} \citep[e.g.,][38]{binmore_natural_2005}: Can we deduct moral principles from reason alone, or do we have to consult data from the real world?
    \item The role of \textbf{intuitions} \citep{gigerenzer_gutfeelings_2007, haidt_righteous_2013}: Are intuitions (gut feelings) the benchmark for ``genuine'' justice or merely heuristics for rational reasoning? 
\end{itemize}

This article focuses on distributive justice, i.e., how people allocate benefits and burdens –- the most important being income for work. 
The most general (and commonly accepted) definition is to give each his or her due, a desert-based approach traced back to Aristotle.
Distributions based on desert (merit) enjoy practical prominence especially in markets \citep{miller_principles_1999, adriaans_equity_2019, richters_making_2021}.
But concerns arise about increasing inequality \citep{piketty_capital_2014}, and distributive justice remains a major societal and social scientific controversy.

Given this diversity, is a single (let alone simple) Theory of Justice at all possible?
\citet[35]{deutsch_distributive_1985} thought not: justice would be complex and variable. 
\citet[sec.~7]{miller_justice_2017} assumed that ``we will need to accept that no comprehensive theory of justice is available to us.''
Phenomenologically, \citet{miller_principles_1999} argued that a plurality of three principles would prevail in modern societies: a (moderate) meritocratic principle in combination with a social principle (need) and political equality.
This ``tripartite approach'' emerged in social psychology in opposition to Equity Theory in the 1970s.
It quickly and consistently found the support of many authors \citep[among them][]{deutsch_equity_1975, leventhal_fairness_1976, lerner_justicemotive_1977, schwinger_threeprinciples_1980, mikula_justice_1980, reis_levels_1986, kabanoff_equity_1991} and today represents one paradigm within the justice discourse of social psychology \citep{lind_study_2020}.
But the systematic relationship between these principles remains unclear, and with his specific pluralist account Miller would only shift the problem from the principles to the social context \citep{honneth_forewordmiller_2008}.

This article aims to develop a coherent picture of when, why and how people allocate benefits and burdens according to merit, need, and equality, setting it in the evolutionary context of reciprocity.
In the terms of \citet{leventhal_beyond_1980}, it is a theory of allocation preference.
It does not discuss non-outcome factors like procedural fairness, for in the end outcomes are decisive. 
Because there has been a shift from the outcome-focused to the relationship-focused paradigm in social psychology, beginning in the middle of the 1970s \citep{lind_study_2020}, many of the works concerning distributive justice cited here date back to the outcome-focused period. 

In Section \ref{justsec_reciprocity} the article depicts the evolutionary roots of reciprocity and how reciprocity translates into the social norm \emph{Meritocratic Principle}. 
Section \ref{justsec_bargain} explains why one justice dimension is not enough. 
Moral reasoning is a communicative strategy for resolving conflict, enhancing status, and maintaining cooperation, making justice rather a social bargain and an optimization problem.
The picture is therefore extended in Section \ref{justsec_extending} by social psychological insights regarding intuitions, rules of thumb, and self-bindings.
Section \ref{justsec_need} and Section \ref{justsec_equality} show how need and equality can be reconciled with merit.
They are auxiliary principles when merit is not effective, not efficient, or not communicable (or simply coincides with equality). 
An important topic is implementation (Section \ref{justsec_implementation}): How is a substantive Theory of Justice ``applied'', especially regarding procedural justice?
Section \ref{justsec_discussion} discusses the results and formulates a conclusion.

\section{Reciprocity as the Foundational Principle of Justice}
\label{justsec_reciprocity}

\subsection{The Sociobiology of Cooperation}
\subsubsection{Selective Forces}
Evolutionary selection has shaped a process in which individuals compete for resources to replicate their genes by reproduction, and humans make no exception.
Besides competition, fitness can also be enhanced by cooperation, but according to the selection rules every individual must benefit from it in the long run. 
Today, the existence of genetic altruism can be ruled out -– this would contradict any evolutionary functional logic \citep[65]{voland_soziobiologie:_2013}. 
Even though human biology does not determine behavior, it limits its flexibility in characteristic ways. 

Among animals, mutualism can be observed regularly (behavior that provides \emph{direct} benefits to every mutualist), often between species \citep{leigh_mutualism_2010}.
But only higher primates seem to have evolved the cognitive abilities required for reciprocity, i.e., accepting costs for the benefit of others which are rewarded later, either by the beneficiaries themselves (direct reciprocity) or by others (indirect reciprocity) \citep{trivers_altruism_1971, trivers_reciprocal_2006}.
Reciprocity is dependent on repeated interactions and it is endangered by cheating, so the cognitive abilities required include individual recognition, temporal discounting, and memory \citep{stevens_evolving_2005}.
In parallel, mechanisms have evolved for altruistic punishment of cheaters, but also for building up reputation and the assignment of social approval (prestige) \citep[74ff.]{voland_soziobiologie:_2013}.

\subsubsection{Forms of Reciprocity}
Reciprocity among humans has been widely discussed in anthropology and sociology \citep[see][]{adloff_vom_2005}, but only recently in psychology \citep{kurzban_evolution_2015}.
Note that some authors, especially in economics, restrict the term reciprocity to a \emph{personal} relation.
Reciprocity implies an \emph{equivalence} in exchange relations: to repay in kind what another has done for us, materially or socially. 
Voluntary exchange will only occur when the benefits exceed the costs for both partners (costs and benefits in a wide sense, \citealt{blau_social_1968, becker_economic_1976}). 
Due to different individual levels of marginal utilities and marginal costs both partners can benefit from exchange.
This proposition lies at the heart of Social Exchange Theory \citep{thibautkelley_social_1959, homans_social_1974, blau_exchange_1964}, and
Neoclassical Economics uses the concept of surplus to characterize such situations.  

The condition of equivalence must be met at least in the long run \citep{kurzban_evolution_2015}.
\citet{sahlins_sociology_1965} introduced the concept of generalized reciprocity: transactions that are seemingly ``altruistic'' but can be expected to be returned, not necessarily here and now, not necessarily by the beneficiaries themselves, undetermined in time, quantity, and value \citep[for a compact overview, see][ch.~5]{holcombe_coordination_2020}.
These transactions might be better described as investments.
The modern welfare state, for example, can be regarded as an institutionally mediated, reciprocal arrangement, and several typologies of reciprocal expectations can be construed \citep{lessenich_reziprozitaet_2005}.
A decisive point is that free-riding is effectively prevented or punished.

\subsubsection{Ideal Justice}
\begin{figure}[t]
\hfil\includegraphics[width=\columnwidth]{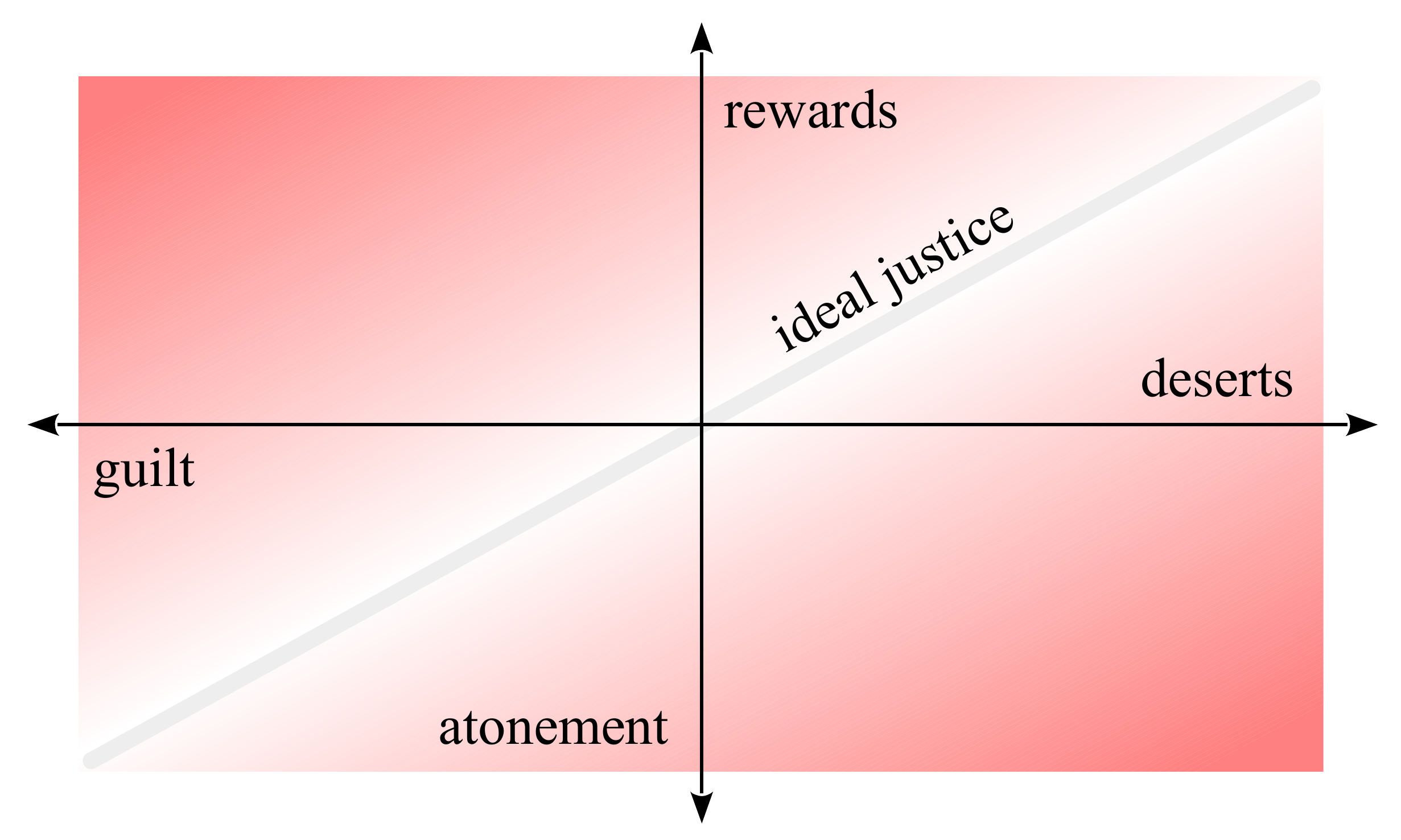}\hfil
    \caption{\label{justfig_idealjustice}
The Ideal of Justice: an exact proportionality between personal costs and personal benefits. For a similar depiction of an equitable relationship, see \citet{walster_correction_1975}.}
\end{figure}

From the evolutionary point of view, equivalence is the only stable (and symmetrical) solution to the problem of how two individuals could possibly engage in voluntary social exchange.
The idea of equivalence can be readily extended to the negative range: The proportional relation between desert and reward also applies to guilt and atonement (Figure \ref{justfig_idealjustice}).
The purpose of punishment is to impose enough costs on the factual or potential defector to offset the temptation to violate the equivalence principle \citep{boyd_punishment_1992, cluttonbrock_punishment_1995}.
\citet[88]{durkheim_division_1933} emphasized the intended \emph{exact} balance between the severity of a crime and its punishment \citep[see also][13]{buchanan_justice_1986}. 
People would even ``construct'' reciprocity if the world is not perceived as just, especially by blaming victims for their ``own fault'' \citep[``Just World Belief''][]{lerner_belief_1980}.
\citet[``Equity with the World'',][]{austin_equity_1975} showed that subjects try to ``heal'' inequities across several relationships, if necessary.

One of the most striking arguments in favor of reciprocity is the existence of property, money, credit, and prices \emph{as such} (\citealt[ch.~5]{holcombe_coordination_2020}).
The idea of property is historically based on the idea of an achievement resulting from one’s own labor \citep{locke_treatises_1967, hume_treatise_1874, smith_inquiry_1981, marx_capital:_1906}, or in a proverb: ``as you sow, so shall you reap''.
An argument often neglected is the proportionality between the amount of money and quantity of goods.
Philosophers of justice regularly argue about the legitimate level of hourly rates \citep[unequal pay, \emph{iustum pretium},][]{koch_okonomische_1995}, but they have a tacit consent that hourly rates \emph{as such} are legitimate, and so are any other prices. 
The equation $amount = quantity \times price$ states that two hours of work will cost two times the amount of one hour \citep[e.g.,][46]{tyler_social_1997}, reflecting the linear relation of Figure \ref{justfig_idealjustice}. 

\subsubsection{Injustices}
The most general definition of a \emph{violation} of reciprocity could be formulated as ``advantages at the expense of others'', or ``not bearing the costs of one’s own choices''.
This holds true for all forms of offense, be it theft, rape, embezzlement, illegal parking, doping or tax evasion.
Other terms indicating unbalanced reciprocal expectations are exploitation, recklessness, usury, plagiarism, cartel, extortion, and slavery.
In economics, \emph{negative externalities} occur when, due to the activity of another person, one person experiences losses in welfare that are not compensated \citep[184]{daly_ecological_2011}.
Under the heading of \emph{rent-seeking}, economists discuss a whole class of activities that attempt to gain larger profits by exploiting or even manipulating economic conditions or public policy, without contributing to genuine wealth creation \citep[e.g.,][]{krueger_political_1974}.
Regulation theory implicitly associates market failures with an imbalance of individual and societal costs and benefits \citep[cf.][449]{shughart_regulation_2008}.
That ``free-riding'' is a social problem remains unintelligible without assuming a violation of reciprocal expectations \citep{fehr_reciprocity_1998, panchanathan_indirect_2004}.

\subsubsection{Formulating the Issue}
\label{justsec_reciprocity_issue}
Therefore, the minimum condition of any kind of voluntary social exchange is reciprocity in the sociobiological notion, a ratio of benefits to (opportunity) costs of unity and one that is not too unequal for both, however this is assessed by the individuals -- and their collectives \citep[cf.][145f.: ``When consent is not enough'']{sandel_justice:_2009}. 
Justice is a social bargain between individuals but witnessed by society. 

\begin{figure}[t]
\hfil\includegraphics[width=\columnwidth]{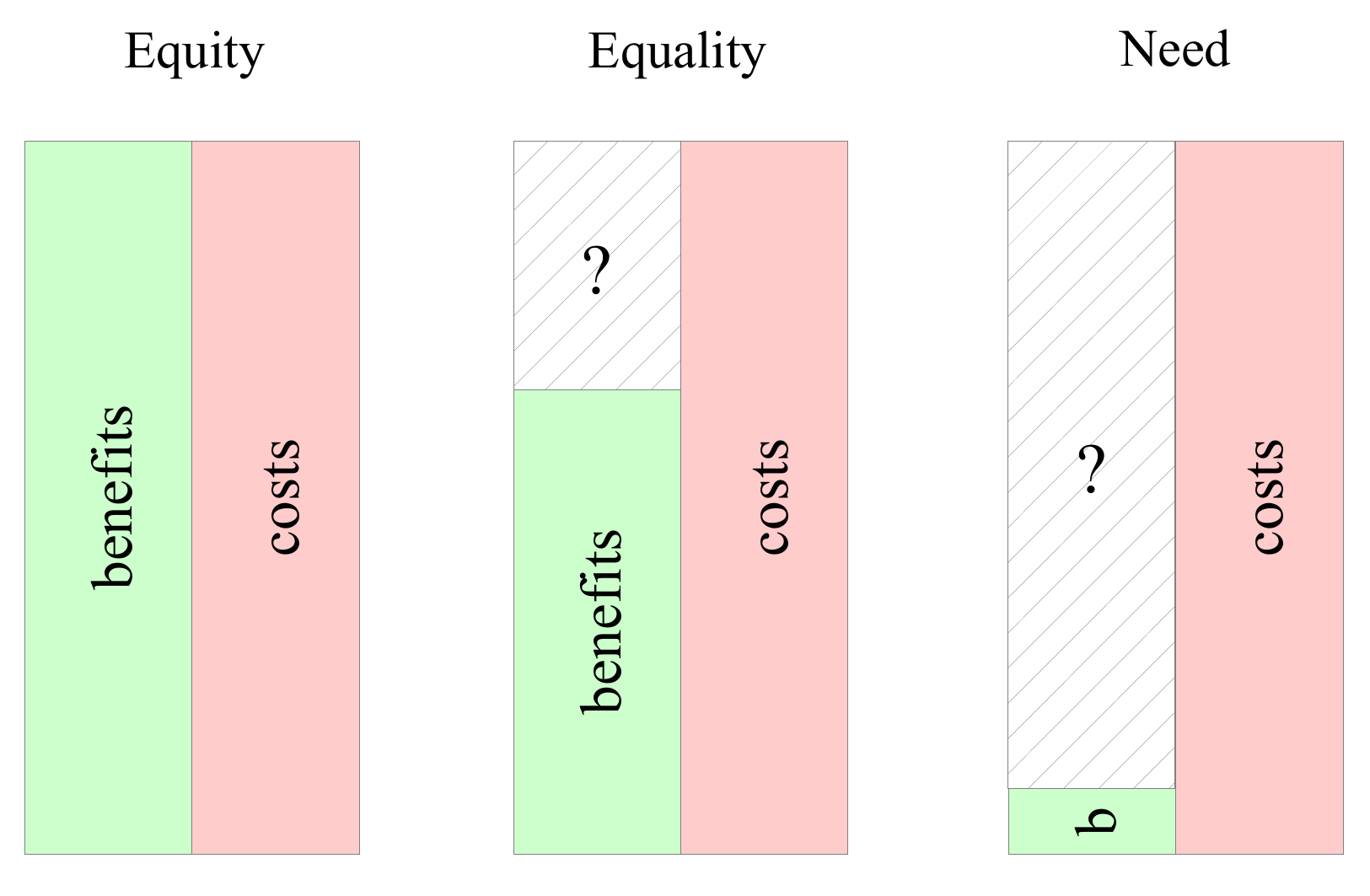}\hfil
    \caption{\label{justfig_gaps}
What could fill the gaps between seemingly higher costs and lower benefits when choosing equality or need as distribution rule?}
\end{figure}

In Theories of Justice, reciprocity is conceptualized as merit, proportionality of contribution, equity, mutuality, or Golden Rule.
The theoretical challenge therefore is to integrate the seemingly evident deviations -- like need, equality, and ``altruistic'' acts -- into a coherent concept of reciprocity: what could fill the gaps between seemingly higher costs and lower benefits in these cases (cf.\ Figure \ref{justfig_gaps})?

It is important not to confuse two overlapping, but distinct assessments of social exchange:
(1) ``Is it just?'' 
This concerns the ``public'' cost-benefit estimation, i.e., the visible (communicable) aspects of reciprocity that are depicted in Figure \ref{justfig_idealjustice}.
(2) ``Is it worth the effort?'' 
This concerns the individual, ``private'' cost-benefit estimation.
Beyond the obvious, it includes costs and benefits that are not well communicable or only indirectly related to the transaction and hence unsure: when do benefits cease to be \emph{related} to a specific social interaction and hence make for ``genuine altruism''?
The difference becomes salient for example when \citet[5]{nesse_natural_2001} objected to ``[t]he tendency to use reciprocity to stand for all cooperative relationships'' and listed several ``other'' reasons why humans would cooperate.

\begin{figure*}[t]
\hfil\includegraphics[width=0.7\textwidth]{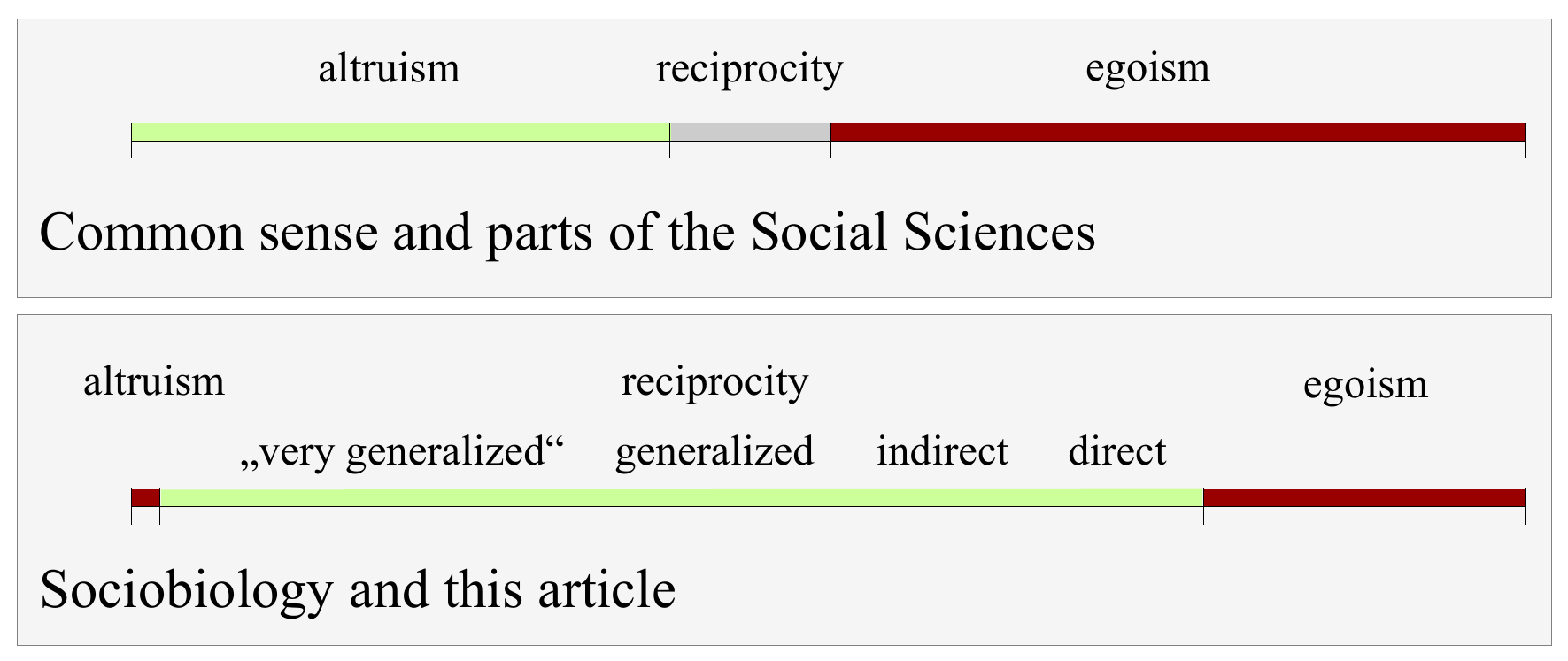}\hfil
    \caption{\label{justfig_reciprocity}
Tensions of Reciprocity between Good and Evil.}
\end{figure*}

Figure \ref{justfig_reciprocity} visualizes some of these tensions. 
While sociobiology concentrates on genetic replication where altruism goes extinct, common sense and parts of the social sciences prefer a semantic dichotomy between ``Good and Evil'', restricting reciprocity to a small range of transparent cases (and sometimes to personal relations). 
The semantics revolve around motives \citep[cf.][ch.~4]{homann_sollen_2014}: ``egoism'' is present when one's own advantages are intended (whether or not at the expense of others), and ``altruism'' is present when behavior is clearly considerate of others (though personal benefits may arise unintendedly). 
This article argues that the domain of reciprocity is much broader and that not intentions, but actual costs and benefits should inform the wording. 
Much semantic egoism is in fact adequate self-regard, and a lot of what is termed altruism should rather be viewed as a ``blind investment''.
Instead of inventing a new term (cf.\ the ``social semantics controversy'', \citealp{west_socialsemantics_2007, west_socialsemantics_2008}, \citealp{wilson_socialsemantics_2008}) I will refer to this as ``very generalized reciprocity''.
True altruism (benefits < costs) as a form of self-exploitation is neither likely nor socially desirable.

\subsection{The Meritocratic Principle}
\label{justsec_reciprocity_principle}

\subsubsection{History of Discourse}
Reciprocity translates into the social norm \emph{Meritocratic Principle}.
The idea of merit is that people \emph{deserve} certain economic benefits in light of their actions \citep{feldman_desert_2016, lamont_distributive_2017}.
Already Aristotle argued in his \emph{Nicomachean Ethics} (1131a) that ``all men agree that what is just in distribution must be according to merit''. 

In Germany, from about 1960 until 1980 three overlapping strands of the debate about ``achievement'' can be distinguished (citations are by way of example):
\begin{itemize}
    \item A wage-political struggle between employers and trade unions that continues to this day (performance-related pay) \citep{pornschlegel_menschliche_1967, verband_fur_arbeitsstudien_refa_leistungsprinzip_1974, breisig_entgelt_2003},
    \item a (leftist) critique of a ``working society'', ``culture of achievement'' and alienation of work -- and the (conservative) reply \citep{offe_leistungsprinzip_1970, muller_sinn_1974, jenkis_leistung_1980}, 
    \item and how social scientists dealt with the subject \citep{heckhausen_leistungchancen_1974, braun_leistung_1977, hartfiel_leistungsprinzip._1977, bolte_leistung_1979}.
\end{itemize}
The debate revived around the year 2000 and culminated in the aftermath of the banking crisis in 2008. 
While the meritocratic principle was fiercely contested in the 1970s, it was now explicitly demanded \citep{neckel_leistung_2001, droge_ruckkehr_2008}.
Most authors mentioned above agree that merit has many dimensions, and that nearly every aspect of merit (or achievement) can be interpreted positively or negatively. 
Any objective definition of merit is impossible since it depends on context and circumstances -- or the other way round: Merit is uncontroversial only when defined in extremely narrow terms (\citealt[cf.][133f.,]{deutsch_distributive_1985}, \citealt[sec.~3.2]{breisig_entgelt_2003}).
Actually, it is always the beneficiaries of an achievement who will assess its utility.
This is a core principle of the division of labor and reflects the general tension between individual and collective perceptions of achievement. 

In the UK and the US, the debate had another focus.
An important precursor was the so-called \textit{Northcote-Trevelyan Report} in Victorian England in 1854. 
Its unanimous recommendation was to organize recruitment and promotion in the British Civil Service exclusively according to merit.
This was partly (and correctly) perceived as a critique of aristocratic patronage.
In his satirical political book ``\textit{The Rise of the Meritocracy}'' \citet{young_rise_1958} coined the term \emph{Meritocracy} as a dystopia, ``portraying a sinister, highly stratified society organised around intelligence testing and educational selection'' \cite[][book cover]{dench_rise_2006}.
A collection published after a symposium \citep{dench_rise_2006} allows an overview of the debate in the UK and the US.

A proportionality of inputs and outcomes was also the core proposition of \textit{Equity Theory} which was the leading paradigm of social psychology during the 1960s and 1970s and led to hopes of becoming a general theory of social interaction \citep{adams_towards_1963, adams_inequity_1965, berkowitz_equitytoward_1976}.
Both inputs and outcomes could be positive or negative, and their relation was presented in a complex, somewhat counterintuitive mathematical formula \citep{walster_directions_1973}.
The theory was later criticized for its ``unidimensionality'' and other conceptual shortcomings and subsequently replaced by the ``tripartite approach'' with merit, need, and equality \citep{leventhal_whatshould_1980, schwinger_threeprinciples_1980, deutsch_distributive_1985, folger_rethinking_1986, reis_levels_1986, lind_study_2020}.

\subsubsection{Definition}
The meritocratic principle has many formulations, but often as simple as this: ``Those who do more than others shall receive more than others'' \citep[26, translated]{bolte_leistung_1979}.
Likewise note that also a ``reversed equity script'' has been reported: ``those who have more must have performed better'' \citep[174]{bierhoff_social_1986}.
A formal (and lengthy) definition can be found in \citet[46]{neckel_umkampfte_2008}, but in my view the best definition of merit (especially for modern societies) is the legendary equation of Michael Young in an extended version \citep[138]{kariya_japan_2006},

\begin{center}
\textit{merit = talent + effort + luck}.
\end{center}

More easily than any other definition it can reveal the wrangling over the role of merit.
This equation also matches economic theory when interpreting its main terms as ``personal production factors'': the natural endowment of individuals with talents (capital) and their efforts (labor), complemented by the contingencies of the economic process and the sometimes hardly predictable assessment by others. 

\subsubsection{Justice \emph{Theory} as a Social Bargain -- Tugging at the Terms of Young's Equation}
Luck is the least contested term in Young’s equation, for it has basically nothing to do with merit.
Nevertheless, the term is important for structural and discursive reasons \citep[cf.][sec.\ 1.7]{droge_ruckkehr_2008}.
First, it is unavoidable, second, people argue about what counts as luck, and third, it is often impossible to clearly separate it from the rest.
\emph{Luck egalitarianism} \citep[e.g.,][]{dworkin_sovereign_2000} is effectively a debate about luck vs.\ choice, i.e., responsibility \citep[cf.][ch.~4]{white_equality_2007}.

At the least, luck must not be too large.
Today, a widespread economic contingency is one of the main acceptancy problems of the meritocratic principle. 
\citet[sec.\ 4]{neckel_leistung_2001} emphasized that merit as a foundation of distribution becomes absurd when jobs themselves have become scarcities (``economic culture of randomness'').
On the other hand, \citet[192]{saunders_meritocracy_2006} argued that luck cannot be systematically exploited, and everyone can have it, so it would not question the legitimacy of meritocracy.

Talent (or ability) is ambiguous.
Exemplary for many, \citet[86]{rawls_theoryrevised_1999} emphasized that ``inequalities of birth and natural endowment are undeserved'', results of a ``natural lottery'' (p.~64).
At the most, achievements based on talent could entitle legitimate expectations from society, given that appropriate rules had been established in advance: ``But this sense of desert is that of entitlement'' (p.~88f.).
Rather like in a tug-of-war, Rawls tried to ``tug'' talent toward the term luck and to turn a moral desert into a question of contingent entitlement. 
Libertarians argued the other way round \citep[cf.][69]{sandel_justice:_2009}: Talent requires a lot of effort to develop (``knowledge''), and who else but the talented should be allowed to benefit from his or her talent?
Libertarians tried to tug talent toward the term effort and turn a moral desert into a question of property rights.
The resemblance to the nature-nurture debate is obvious. 

The term effort is indeed for many the allegory of a deserved income. 
Everywhere in the world hard working people are acknowledged as creators of economic value -- at least if this effort is targeted at a socially desirable outcome and achieves a result. 
Throughout the debate the tension between effort and achievement is a recurring topic: effort without achievement is a pity, achievement without effort has no merit \citep{heckhausen_leistungchancen_1974, braun_leistung_1977, bolte_leistung_1979, deutsch_distributive_1985, neckel_leistung_2001, droge_ruckkehr_2008}.  
Neither entitles reward, and it could even be argued that the equation should rather read merit = talent $\times$ effort + luck, so that if either talent or effort are zero, their product would be zero as well, the remainder being luck. 
Yet the basic problem of ``performance-related pay'' is that while its \emph{normative} basis can only be merit, its \emph{transparent} (communicable) basis can only be achievements \citep[sec.~2.2]{breisig_entgelt_2003}. 

\citet[sec.~12]{rawls_theoryrevised_1999} went even further and declared that the ability to make an effort was dependent upon a happy childhood and social circumstances \citep[cf.\ also][158]{sandel_justice:_2009}. 
In fact, Rawls tugged \emph{any} term of Young’s equation toward luck, apparently to be free for his contractarian approach to justice, i.e., contingent agreement \citep[cf.][pp.~53-59]{miller_principles_1999}.

Karl Marx concentrated on effort and called this, following his classical antecessors, a \emph{labor} theory of value. 
In his writings Marx spoke of the equality of human work, of the time elapsed as a measure of effort \citep[71ff.]{heinrich_kritik_2005}, but also that a higher qualification would cause higher costs, and hence justify a higher value of the working time (hourly rate, p.~91). 
He largely ignored talent and fully disregarded the contribution of a whole class, namely capital as the merit of the entrepreneurs, which he assumed to be mere organization plus supervision of exploitation and therefore zero (p.~157).

Talent and effort have yet another dimension: Societal functionality. 
Early accounts of the social benefits resulting from functional stratification are \citet{durkheim_division_1933} and \citet{davis_stratification_1945}.
The importance of not discouraging the talented is another recurring topic of the debate about merit, tightly connected to secure property rights \citep[cf.][]{acemoglu_nations_2012}.

\subsubsection{Conclusion}
Contrary to Young's apprehension and other objections, the meritocratic principle is a fundamental and widely accepted social norm:
Meritocracy ``resonates powerfully with deeply held ethical values about \emph{fairness}, and these are broadly shared throughout the population'' \citep[193, original emphasis]{saunders_meritocracy_2006}.
It ``corresponds to the widespread belief that people deserve to enjoy unequal incomes depending on their abilities and how hard they work'' \citep[178]{miller_principles_1999}.
The meritocratic principle establishes a relationship between personal market value and contribution to productivity \citep[159]{marris_just_2006}.

Unlike many other theorists, \citet{miller_principles_1999} considered popular conceptions of justice (``what the people think'') and empirical data.
He stated that desert and need criteria feature prominently here (p.~90), and criticized political philosophers, Rawls among them, for neglecting such empirical evidence, especially when it is in favor of desert (ch.\ 7).
Further empirical support for merit as a base for distributive justice can be found in \citet{saunders_meritocracy_2006} and \citet{neckel_umkampfte_2008}.
Recently, \citet{adriaans_equity_2019} showed that there is widespread approval in Germany and the other European countries for income distribution according to equity and need.

The results of the scientific and philosophical debates about the meritocratic principle can be summarized as follows:
\begin{itemize}
    \item It is defendable as a theoretical base of justice.
    \item It is functional in increasing societal wealth.
    \item It is not possible to define merit objectively.
    \item In practice, the principle is often not met (effortless incomes, underpaid work, the competing principles need and equality).
\end{itemize}

Most objections against the meritocratic principle can be rebutted by concentrating on ``non-merit'' (Section \ref{justsec_implementation}).

\section{Justice as a Social Bargain and Optimization Problem}
\label{justsec_bargain}

Figure \ref{justfig_idealjustice} depicts the ideal of justice, demanding to appreciate the circumstances (costs and benefits) of every single case.
But neither our physical nor our social world are ideal. 
Since justice is a discursive, nonviolent concept \citep[like arguments in general,][]{kopperschmidt_argumentationstheorie_2000}, we can at least set aside questions of power and force for a \emph{Theory} of Justice.
Then still among the real-world restrictions are (1)~costs, time constraints and scope for information retrieval and communication, (2)~general uncertainty about future events, (3)~strategic behavior of our fellow humans.

Moral reasoning is a communicative strategy for resolving conflict, enhancing status, and maintaining cooperation (\citealt[cf.][26]{deutsch_distributive_1985}, \citealt{mikula_justice_2000}, \citealt{haidt_righteous_2013}), making justice rather a social bargain and an optimization problem.
It includes assessments, compromises, investments, renunciation -- and above all a pool of shared ``good reasons'' (plausibility resources) as an argumentative basis \citep{kopperschmidt_argumentationstheorie_2000}.

The ideal of justice is just one of many dimensions (\citealt[cf.][]{reis_multidimensionality_1984}, \citealt[99]{deutsch_distributive_1985}, \citealt{young_equity_1994}), and not always the strongest force. 
In practice, justice must in my view fulfill at least three additional requirements: effectiveness, efficiency, and communicability.
For each the Aristotelian ``mean between extremes'' is optimal, but since trade-offs exist between them, we actually have an optimization problem.

\textbf{Effectiveness}: functionalist $\sim$ expedient $\sim$ sticking to principles

Justice must be useful, that is the quintessence of utilitarianism. 
If justice degenerates into ``pure principle'' (dogmatism), creating problems rather than solving them, it misses the target (e.g., Kant’s murderer dilemma, \citealt[132]{sandel_justice:_2009}).
\citet{mikula_justice_1980} discusses extensively the role of ``allocation goals'', and \citet{greenberg_approaching_1982} and \citet{leventhal_distribution_1976} are rich sources of pragmatic considerations when choosing an allocation rule. 
Effectiveness has priority when it comes to rationing \citep{elster_local_1992, young_equity_1994}. 
On the other hand, justice must not be functionalist, that is the quintessence of deontology. 
If justice uncritically maximizes ``outcome'' (especially in the short run) it becomes arbitrary, and many individual and societal self-bindings remain unintelligible \citep[32f.]{sandel_justice:_2009}.

\textbf{Efficiency}: sweeping $\sim$ practicable $\sim$ sophisticated

Since humans have evolved to maximize efficiency \citep[148]{sanderson_evolution_2001}, justice should be rather straightforward and ``easy'', at least practicable. 
It must be applicable to everyday situations, not only to philosophical thought experiments.
Long theories, complex reasoning and the elaborate dissection of moral subtleties are as ``unjust'' as endless meters of tax legislation and its commentaries. 
But justice cannot be allowed to simply flatten all relevant differences between the cases in question in order to push through one single principle. 
We cannot deviate too far from our ideal of justice to appreciate the individual circumstances of every single case.

\textbf{Communicability}: populist $\sim$ thoughtful $\sim$ unworldly

Justice must be ``reasonable''. 
A Theory of Justice must be consistent and withstand rational reflection. 
Any distribution rule must fulfill the rule of the good reason (Section \ref{justsec_extending_decision}) and ``treat like cases alike''.
Armchair theories that contradict basic intuitions of justice and ignore empirical evidence are to be refuted \citep{miller_principles_1999}. 
On the other hand, justice cannot be merely a question of majority (\emph{vox populi}) or strong emotions.
A precondition of a shared sense of justice is comprehensibility and transparency of the criteria. 
Justice must be socially monitorable. 

To derive the proper consequences for a Theory of Justice we must extend the picture, because we need to know more about the social psychology of humans regarding (moral) intuitions, the role of rules of thumb, good reasons, and simplicity in efficient decision making, and -- especially important -- self-bindings as a precautionary device and ``blind investment''.

\section{Extending the Picture}
\label{justsec_extending}

\subsection{Self-Binding}
\label{justsec_extending_selfbinding}

\subsubsection{Theorists of Self-Binding}

Self-binding (also called commitment or precommitment) was made a topic in economics by Thomas C.\ \citet{schelling_strategy_1960}.
He regarded conflict situations between several parties rather as \emph{bargaining situations}, and believed that the public and credible renunciation of realistic options could improve one’s bargaining position.
This was in stark contrast to the neoclassical assumption that more options are always better. 
``Burning one's bridges behind oneself'' to unmistakably make clear that retreat is no option, or ``calculated madness'' (brinkmanship) can force the other parties to make a complete reassessment of the whole situation.
Schelling later extended the idea of conflict to intra-personal ``two-mindedness'' \citep{schelling_egonomics_1978, schelling_selfcommand_1984, schelling_enforcing_1985}.

The idea of commitment can be readily extended from conflict situations to social behavior in general, a topic of Jon Elster's (e.g., \citeyear{elster_ulysses_1979}).
A self-binding is to act now to ensure a future act that \emph{could} but would not have been performed without that prior act \citep[1754, discussing constitutions as collective self-bindings]{elster_dontburn_2003}.
Since by definition trade-offs exist, any self-binding is endangered by a change of mind, therefore in Elster's opinion a self-binding must be public, and its revocation must be costly (or impossible). 
Elster is very strict about two assumptions: 
(1)~The outcome must be intended. 
Any ``unconscious'' self-binding would devalue it.
(2)~Only individuals can bind themselves. 
Any notion of collective self-binding (institutions) is to be explained in terms of individual (\emph{and} intentional) self-binding.
Elster makes further comparisons between individual commitment and constitutions: Despite structural similarities, an individual would lack the possibility of a separation of powers, and attaching costs to some options would be no feature of constitutions \citep[1754f.]{elster_dontburn_2003}.

For Karl \citet[part B]{homann_2003_anreize}, self-binding is a means for more freedom.
Self-bindings make individuals \emph{reliable} by removing unfair options from their repertoire of actions, and this reliability of mutual expectations enables new and beneficial cooperation between self-interested individuals (p.~73 and ch.\ 6). 
Homann is a critic of ``too much'' individual self-binding (decency, appeals to values, Christian ethics etc.), and he accuses philosophical ethics of being heavily biased toward the idea of a ``genuinely moral motive'' (cf.\ Section \ref{justsec_implementation}).
Most modern problems would instead be the non-intended side effects of ``normal'' behavior, so any higher morality would be no remedy, especially because of the exploitability by others. 
Instead, he emphasizes the significance of institutions as collective self-bindings.

\subsubsection{Properties and Examples of Self-Bindings}
Self-bindings are part of the ``little tricks we play on ourselves'' \citep[290]{schelling_egonomics_1978} to limit our potential to dupe ourselves.
This is a classic case of utility maximization in the \emph{long run} (the time scale considered is crucial in this respect).
For the topics discussed here I suggest the following definition:

\begin{quote}
A self-binding is the strict adherence to a contingent rule which aims at some ``higher goal'' that is difficult to achieve directly. 
\end{quote}

I prefer the term self-binding over commitment because the latter is too near to a (personal) promise and lacks the contingency aspect.
A self-binding lies somewhere between ``pure investment'' where costs now will surely reap benefits later, and ``pure superstition'' where costs now definitely have no connection to assumed benefits later. 
``Higher goal'' means that the connection between the rule (costs) and the desired outcome (benefits) is not transparent.
Their probabilistic relation may be plausible in some cases but controversial in others (see examples below). 
Therefore, the rule is contingent.
But the rule must be strictly adhered to for it to be communicable, fulfillable (clear criteria), and socially monitorable. 
Hence, a self-binding could also be viewed as a ``blind investment''.
In my view it is speculations about the long-term significance (``sense'') of certain self-bindings (``values'') and the individual (non)willingness to make such blind investments that make societal debates so wide-ranging (and indeed heated).

I would rather give up several of Elster’s restrictions:
\begin{itemize}
    \item There is no need to restrict self-bindings to be intentional (conscious). 
    Many interesting self-bindings would not pass this test, and even Elster struggles with that somewhat contingent distinction \citep[e.g.,][1761]{elster_dontburn_2003}. 
    The unconscious is \emph{our} unconscious that pursues \emph{our} goals and can make judgements on its own \citep[275, original emphasis]{sumser_evolution_2016}.
    \item A self-binding does not need to be public. 
    Any intent (e.g., not becoming addicted to alcohol) that is safeguarded by a contingent rule (``no alcohol before 8 p.m.'') exhibits the characteristics of a self-binding, a view supported by Schelling throughout his contributions.
    \item Contrary to Elster, personal self-bindings could be effective \emph{because} individuals have a separation of powers, identifying the roles of individual action, moral standards, and conscience (guilt, shame) with the executive, legislative and judiciary, as elements of intrapersonal communication \citep[cf.\ the theory of self-command,][]{schelling_egonomics_1978}. 
\end{itemize}

Self-bindings can be applied on different levels, from the individual up to the global community, and their characteristics depend on goal, criticality (what is at stake) and transparency (communicability).
A structural characteristic is that a self-binding on a higher level (collective) is a coercion on the lower level (the collective’s individuals) when a self-binding on the individual level seems rather unpromising. 

The following list of examples is neither exhaustive nor strict in its categories \citep[see also][, for an overview]{nesse_evolution_2001}:

\begin{itemize}
    \item On the \textbf{individual level}, we find concepts like civility and good conduct, moral principles, loyalty, veganism or a ``pro-life'' stance, awareness-building, religious dietary laws, superstition, and political correctness. 
    People \emph{believe} rather than know that certain behavior would be favorable or adverse for themselves or for running a society, and they discuss passionately about it \citep{haidt_righteous_2013}.
    Self-bindings can be part of a (political) identity and contribute to authenticity \citep[e.g.,][]{greenebaum_veganism_2012}.
    The higher goal of credibility may sometimes require even ``silly'' investments \cite[keeping a ruinous promise, carrying out a senseless threat,][12]{nesse_natural_2001}.
    \item On the level of \textbf{societal subsystems}, we have most prominently science and the judiciary.
    For the higher goal of finding ``objective truth'' in a society of strategic self-interested individuals, both have given their systems strict, transparent rules so as to avoid or correct any biases of their members and maintain a high standard of impartiality.
    These rules are continuously reflected upon.
    \item On the level of \textbf{societies} (nation states), we observe restrictions regarding the ownership of weapons or consumption of drugs. 
    During the Corona pandemic many legislations have created a host of contingent, quickly changing and hence contested rules regarding personal contacts. 
    Constitutions are among the highest-level national self-bindings we know, ``intricate pieces of machinery'' \citep[1779]{elster_dontburn_2003}.
    Between nations we find concepts like nuclear deterrence or Swiss and Swedish neutrality.
    \item On the level of \textbf{God}, the existence or lack of self-binding has been discussed since time immemorial (``Theodicy''). 
\end{itemize}

Self-bindings can be nested (several levels of judicial appeal), staggered (adherence to a self-binding is safeguarded by another self-binding), and distributed (network of mutual control, like the ``scientific community'' or separation of powers).

% Critical self-bindings must have huge safety margins which make them disputable.
% A severe work accident in an industrial plant is at the top of a pyramid, with many smaller incidents in the middle and even more situations at the base that are ``only'' critical. 
% ``Tackling the accident pyramid at its bottom means cutting back its top'' (a CSO, quoted in \citealt[179]{gottschall_management_1992}).
% This entails a whole system of rather simple precautionary measures that increase safety but reduce efficiency.
% But often a disaster is the consequence of many small neglects, and the ``Aristotelian'' approach of appreciating the circumstances of every situation is not efficient either. 
% In the long run, ``reasonable'' self-bindings are optimal solutions. 

The tension between rigidity and flexibility of self-bindings is a recurrent topic.
The ``dogmatic'' adherence to a self-binding is in fact its caricature (because the rule itself is contingent), but a ``flexible'' self-binding is neither reliable nor communicable or monitorable, so at least public collective self-bindings can be nothing but ``rigid''.
Handling self-bindings inconsistently foments a suspicion of bias. 
Practically, when a self-binding becomes absurd (``Fiat iustitia et pereat mundus'') or when all people concerned share a strong interest in not following the rule, they usually search for justifications to discard it or to go to the limits of latitude.
% But this remains dangerous: ``In practice, we may not be able to draw the line between temptations and legitimate exceptions'' \citep[1787]{elster_dontburn_2003}.

\subsection{Moral Intuitions}
\label{justsec_extending_moralintuitions}

Gut feelings (intuitions, hunches) refer to a judgement (1)~that appears quickly in our consciousness, (2)~whose underlying reasons we are not fully aware of, and (3)~that is strong enough to act upon \citep[16ff.]{gigerenzer_gutfeelings_2007}.
Intuitions work by using simple rules of thumb (heuristics), which take advantage of evolved capacities of the brain.
Although intuitions often use very little information, they can be amazingly accurate and better (or at least not worse) than much more complex strategies (ch.~8).
The ``intelligence of the unconscious'' lies in choosing a rule of thumb that is appropriate for the situation and relying only on the most relevant information (``the essentials'') while neglecting the rest.

Humans are social, and humankind has evolved much of their time in groups of manageable size. 
The complementarity of individual development and group stability requires a delicate balance (or trade-off) between striving for autonomy and accepting heteronomy and hierarchy. 
Moral intuitions assist in finding that individual balance \citep[cf.][sec.~4.2]{sumser_evolution_2016}.
Although moral rules seem to hold universally, the subject experiences the source of its moral judgement in itself, connected with sentiments of obligation and commitment.
Regarding morals, everyone is an expert. 

In the last two decades, two models developed for understanding moral judgement have garnered prominence.
\citet{haidt_emotionaldog_2001} presented the ``Social Intuitionist Model'' (SIM) which initiated a turn from rationalism to intuitionism in moral psychology \citep[cf.\ also][]{haidt_righteous_2013}.
According to the SIM, rational reflection is neither the cause nor even the starting point of a moral judgement.
Instead, ``intuitions come first, strategic reasoning second'' \citep[xiv]{haidt_righteous_2013}.
Moral intuitions have a primacy, they allow an immediate and effortless assessment of moral situations. 
The \textit{post hoc} reasoning has above all a communicative function of (social) justification. 
Reasons do not have to be true but rather plausible and acceptable (``press secretary, not independent judge''). 
Intuitions are not immune to the reflection of ``thoughtful and well-educated'' people \citep[40]{ross_right_1930}, %cited acc.\ to \citealt[279]{audi_moral_2002}),
but to change one's mind they usually need to activate another intuition. 
For justice, especially significant is the tension between a strongly felt moral conviction and its bad communicability \citep[``moral dumbfounding'',][ch.~2]{haidt_righteous_2013}.

The SIM was challenged by ``Triune Ethics Theory'' (TET, \citealt{narvaez_triune_2008, narvaez_moral_2010}).
According to TET, reflection plays a much larger role than suggested by the SIM.
Reflection can be made possible by conscious deliberation, moral development, and developing expertise.
These abilities can be somatized again and then ``look like intuition'' \citep[167]{narvaez_moral_2010}.

Both models can be viewed as quite complementary rather than contrary, the SIM describing more everyday situations, the TET more ``professional'' moral judgements. 
This way or the other, moral considerations begin with an intuition, may be consciously reflected if necessary, and become a judgement only when it feels emotionally safe \citep[sec.~8.1]{sumser_evolution_2016}. 
Social framing can strongly influence moral judgements, as can priming \cite[emotional flashes like disgust,][]{haidt_righteous_2013}.

\subsection{Efficient Decision-Making}
\label{justsec_extending_decision}

\subsubsection{(No) Good Reason}
According to \citet[ch.~8 and references therein]{gigerenzer_gutfeelings_2007}, our intuitive judgements are often based on only \emph{one good reason}.
This is contrary to the tenets of rational decision theory which holds that we must consider and weigh all relevant information. 
A reason is a cue or signal that is significant for a relevant fact, and humans use reasons to make decisions intuitively and efficiently.  
The Take-the-Best heuristic is an application of that principle: reasons are ordered by relevance and then checked successively (lexicographically) as to whether they enable a decision. 
Instead of weighing all options ``until the end'', the most relevant reasons are checked first. 

\subsubsection{Rules of Thumb}
A heuristic is a problem-solving strategy for \emph{quickly} finding a \emph{satisfactory} solution at the expense of accuracy, especially when the problem is complex or when information is scarce.
A rule of thumb can be defined as a \emph{default} heuristic for a \emph{specific} problem that is regularly encountered. 
Rules of thumb are applied in bodily movements (e.g., catching a flying ball) as well as in mental processes. 
They play an important role in moral behavior, the paradigmatic example being the \emph{Golden Rule}: ``Treat others as you would like others to treat you''.  
Another prominent example for a rule of thumb is ``tit for tat'', a simple recipe for successful cooperation without being exploited \citep{axelrod_evolution_1981}: Start with cooperation and then replicate every opponent's move (cooperation or defection). 
Both rules are strictly reciprocal. 

\citet[ch.~10]{gigerenzer_gutfeelings_2007} states two examples where rules of thumb can be viewed as \emph{social instincts}:
(1)~Strong group cohesion supports conformity to group norms and hierarchy, and the rule ``don't break ranks'' can explain cases of participation even in barbaric acts against one's own convictions, as for example in war. 
(2)~Many jurisdictions require potential donors of organs to consent in advance, but the degree of consent depends heavily on the design of the questionnaire (\emph{nudging}): Most people choose the ``default option'', i.e., the option that for some reason is more straightforward than the others or even pre-selected \citep[``focal point'',][]{schelling_strategy_1960}. 
The rule of thumb reads: ``If there is a default, do nothing about it''.
Another example given for social instincts is imitation \citep[ch.~11]{gigerenzer_gutfeelings_2007}: ``Do what the majority of your peers do'' is a good recipe for avoiding failure, but it can be overruled by ``Do what the successful do''.

Rules of thumb are applied quickly, effortlessly, and usually unconsciously, but can be -- as suggested by SIM and TET -- modified or overruled by reflection, though this is rare. 
Another rule of thumb especially important for a Theory of Justice -- equality as default -- is discussed in Section \ref{justsec_equality_default}.

\section{Need as Auxiliary Principle}
\label{justsec_need}

\subsection{The Domain of Need}
Of all principles, need seems to be most apt to explain ``altruism'': people not able to maintain themselves receive aid from others who cannot expect any reciprocal return \citep[``giving something for nothing'',][]{gouldner_something_1973}. 
But \citet[582]{kurzban_evolution_2015} pointed to the fact that the costs for any aid for non-kin must be recovered at least on average, otherwise this behavior could not have been selected for in the evolutionary process --  hence, it is not altruistic.

This section discusses two topics:  
(1)~Exchange in the family is governed by reciprocity, not need. 
(2)~Institutionalized aid to non-kin (welfare) is dependent on many preconditions, the most important being ``deservingness'', what can be best interpreted in terms of reciprocal expectations.

\subsection{Need in the Family?}
\label{justsec_need_family}
Frequently, families are stated as a counter example to reciprocity, since children would not be able to reciprocate (\citealt{gouldner_something_1973}, \citealt[51]{tyler_social_1997}, \citealt[130]{heidenreich_theorien_2011}), or families would obviously distribute mainly according to equality and/or need (\citealt[153]{lerner_deserving_1976}, \citealt[107]{hochschild_what_1981}, \citealt[pp.~29f., 42f.,]{deutsch_distributive_1985}, \citealt[26]{miller_principles_1999}).
Viewed sociobiologically, it is evident that kin selection rules family life, but a reciprocal vocabulary can also be used.

The basic ``utility'' of children lies in replicating their parents' genes, and evolution could hardly select for anything else. 
\citet[ch.~6]{walster_equity_1978} discuss in depth the costs and rewards of the parent-child relationship in terms of Equity Theory, and they cite evidence that a kind of ``immortality'' is a strong source of parental satisfaction. 
It is part of the parental dilemma that children accomplish this ``achievement'' simply by birth, and from that moment on their genetic interests differ from that of their parents \citep{kurzban_evolution_2015}.
These genetic parent-child conflicts become evident for example in weaning conflicts and cry-babies \citep[sec.~4.6]{voland_soziobiologie:_2013}. 
If children do not ``cooperate'', parents cannot withdraw their support without risking their parental investment, and parental love has evolved to mitigate this dilemma \citep{trivers_parental_1972, daly_cinderella_1999}.

This ``economic'' or ``biologistic'' wording may sound repulsive \citep[cf.][91ff., wrestling with this question]{graeber_debt:_2011}, but sociobiology provides overwhelming evidence that it is the \emph{biological} descendancy that is decisive for parental investment, not the ``family'' social unit (for the following cf.\ \citealt[ch.~3 \& 4]{voland_soziobiologie:_2013}, with even more examples, or \citealt{daly_cinderella_1999}):
\begin{itemize}
    \item Frustrated reproductive expectations are probably a main reason for divorces, and in many cultures any ``constraints'' on reproductive performance are accepted as reasons for divorce. %Voland S. 139
    Stepchildren, unsure biological fatherhood, and sick or disabled children are destabilizing factors for marriage. 
    \item Stepchildren or children with unsure biological fatherhood suffer massively higher risks of accident and death, have higher stress levels and are often materially disadvantaged, compared to biological children.
    \item Parents try to massively intervene in decisions of their children, ignoring their ``needs'', e.g., with mating prohibitions (``Romeo and Juliet'') or appeals to a ``family solidarity''.
\end{itemize}

Although ``Life Course Reciprocity'' plays an important role for both directions of intergenerational exchange \citep{silverstein_reciprocity_2002}, parents do more for their children than vice versa during their lifespan, even when \emph{not} considering childhood and youth phase \citep[196f.]{hollstein_reziprozitaet_2005}.
All this makes clear that in the family ``need'' is rather a small wave on a deep lake of (hardly quantifiable) reciprocity, and many (grand) parental responses to the needs of their (grand) children can be rightfully regarded as investments.

\subsection{The Role of Deservingness for Welfare}
Modern societies have usually institutionalized aid to non-kin as ``welfare'', with full social inclusion as one of the central aims of social policy \citep[see][, for an overview]{vanOorschot_social_2017}. 
This is not self-evident since there is ``a social norm against living off other people and a corresponding normative pressure to earn one's income from work'' \citep[101]{elster_social_1989}.

During the last decades there has been a growing body of literature on the role of ``desert'' in social welfare, because ``the deservingness opinions of various social actors play a pivotal role in the social legitimacy of welfare schemes'' \citep[4]{vanOorschot_social_2017}. 
Especially important is the repeated finding of a ``universal dimension of support'' for certain welfare schemes: ``[T]he rank order of the average deservingness of the groups of `the elderly', `the sick and disabled', `the unemployed' and `immigrants' tends to be the same'' in all European countries \citep[cf.][14, 20f., and references therein]{vanOorschot_social_2017}.
The basic conditions for the legitimacy of welfare are that (1)~aid as such is restricted to those who ``deserve'' it, and (2)~the extent of aid is restricted to ``need'', i.e., objective and legitimate necessities.

\citet{vanOorschot_who_2000} has developed a framework of five criteria (the ``CARIN'' scheme) that can largely explain differences of the perceived legitimacy of targeting aid to target groups. 
Even though reciprocity is among the criteria, this is ``obvious'' reciprocity.
But all criteria can serve as proxies for justifying a ``very generalized reciprocity'' on the part of the donors, and they match with the ideal of justice. 
They exhibit the properties of good reasons and are well communicable. 

\begin{itemize}
    \item \emph{Control}: Fate or fault?
    If the recipients are personally responsible for their predicament, they are less deserving (if at all).
    Regarding unemployment, an important parameter is the availability of jobs: A high unemployment rate reduces the personal responsibility for being jobless \citep{fridberg_public_2000, jeene_dynamics_2014}. 
    My interpretation: Violating the norm of personal responsibility reveals a lack of ``reciprocal precaution'' – or in other words: guilt, and the ideal of justice requires some atonement to prevent thoughtlessness from being an attractive option.
    \item \emph{Attitude}: Eager or sloppy? 
    Signs of compliance to welfare conditions and to expectations of the donors in general (e.g., gratitude) increase deservingness.
    \citet{kootstra_deserving_2016} found that receivers of welfare who have a long work history and invest great efforts into finding a new job are considered more deserving.
    My interpretation: These signs are proxies for social reliability, a will to engage in one's own contributions and not exploit the situation. 
    \item \emph{Reciprocity}: Gift or compensation?
    During the need situation, reciprocity cannot be expected, but already acquired merits (work history) as well as merits likely in the future increase deservingness, for obvious reasons. 
    \item \emph{Identity}: Us or them?
    The ``closer'' the recipients are to the donors, the higher the perceived deservingness is. 
    My interpretation: The paradigmatic example of closeness regarding welfare is ``nation'', and \citet{miller_nationality_1995} provided an account of how a ``well understood'' sense of nationality can contribute to adhering to a nation as an ethic community. 
    A functioning nation is a good precondition for a high level of generalized reciprocity, as opposed to mere fairness, since generalized reciprocity affords a limited group (p.~70ff.).
    \item \emph{Need}: Basic or superfluous?
    Greater need means more deservingness.
    My interpretation: It is always the donors who define what is considered as need, and they restrict aid to what is necessary, i.e., ``true'' or ``objective'' needs. 
    Any kind of institutionalized welfare aid has relatively low upper limits (``socio-cultural breadline''), while the avoidance of misery specifies the lower limit of aid.
    Extensive checks of available means of the needy are common.
    This can be understood as limiting the deviations from reciprocity to an acceptable extent.
\end{itemize}

\subsection{Summary}
The notion of desert governs the domain of need, and desert is used to justify reciprocal expectations and deviations from it. 
Need is no desert, but it can (and should) be a case for entitlement.
Need has already been interpreted as a metaphor for a Mutual Insurance Society \citep{lucas_justice_1972}, the main fear being free-riding, so several precautions and limitations are necessary. 
Acts of welfare can be interpreted as a ``very generalized reciprocity'', an investment of people into the cohesion of society -- \emph{their} society.

\section{Equality as Auxiliary Principle}
\label{justsec_equality}

\subsection{The Domain of Equality}
\label{justsec_equality_domain}

As a principle of distributive justice, equality is much more demanding than need since need is quite well-defined and has a clear scope. 
Equality is less specific, and especially for distributing income it is of little empirical relevance (see below). 

This section discusses four topics: 
(1)~Equality can be efficient as a default (or fallback) when lacking good reasons to choose otherwise.
(2)~Equality of material rewards for individual achievements (income) is restricted to groups of manageable size and can be reconciled with reciprocity.
(3)~The access to a decision about the distribution of (dis)advantages is equal for all (\emph{equality of access}).
But the decision itself is basically about merit or guilt, and equality of access can be best interpreted as a self-binding. 
(4)~\emph{Political equality} is primarily the lack of good reasons for inequality (and a self-binding). 

\subsection{Equality as Default}
\label{justsec_equality_default}

Equality can be a fallback when the relevance of other criteria is not given or not known precisely enough.
``With a good reason, accept inequality -- with no good reason, strive for equality'' is a widely cited rule of thumb (\citealt[305]{berlin_equality_1956}, \citealt[5]{frankena_concept_1962}, \citealt[128]{benn_principles_1965}; indirectly \citealt[65]{schelling_strategy_1960}, \citealt[140]{mikula_ontherole_1980}, \citealt[15]{buchanan_justice_1986}, \citealt[70]{elster_local_1992}, \citealt[8]{young_equity_1994}, \citealt[233]{miller_principles_1999}, \citealt[11]{white_equality_2007}, \citealt[177]{heidenreich_theorien_2011}; experimentally \citealt[ch.~11]{deutsch_distributive_1985}; dissenting: \citealt[37]{kolm_modern_1996}; for even more references see \citealt[sec.~2.4]{gosepath_equality_2011}).
Generally, justice theory can be stated as \emph{justifying deviations from equality} \citep[200, original emphasis]{elster_local_1992}.

When lacking a good reason, equality is always communicable and difficult to challenge, it ``fills the vacuum of indeterminacy'' \citep[73]{schelling_strategy_1960}.
An equal distribution has the additional advantage of not requiring any effort to assess individual contributions and to defend the final distribution \citep[131]{mikula_ontherole_1980} -- equality is easy, and \citet[213]{walster_equity_1978} provide some efficiency considerations regarding ``When Equality? When Proportionality?'': time constraints, communication costs, the value in dispute, or the significance (precedence) for future decisions. 
But regarding communicability, even the proportionality rule is ``equality'', albeit of \emph{units of claim} \citep[80]{young_equity_1994}.

\subsection{Equality in Groups}
\label{justsec_equality_groups}

\subsubsection{Overview}

Since the 1970s, and in contrast to Equity Theory, several social psychologists have developed multi-dimensional approaches to justice. 
They connected distributive principles with types of social relationships in groups and their specific goals, characterized by a tension between equality and equity \citep[cf.][389f.]{druckman_justice_2016}. 
For example, \citet{deutsch_distributive_1985} contrasted an egalitarian, solidarity-oriented group (focus on social relations) with a meritocratic, economic-oriented system (focus on individual outcomes).
\citet{kabanoff_equity_1991} observed equality combined with an emphasis on solidarity, while equity (proportionality) was combined with an emphasis on productivity. 
%A preference of equality even in ``meritorious'' situations would characterize positive relationships or liking \citep{mikula_sympathie_1973}, long-term relationships \citep{mikula_justice_1980}, and an emphasis on the team aspect of a relationship \citep{lerner_justice_1974}.

Solidarity should not be confused with the principle of need.
Solidarity is a \emph{reciprocal} relation with a norm of mutual obligation to contribute one's share to a common goal \citep{bayertz_four_1999}. %esp. [18f.]
Group dynamics show the interdependence of group cohesiveness, performance, and task commitment \citep{mullen_relation_1994, porter_teamperformance_1996}.

\subsubsection{Confluence of Equity and Equality}
Distributions based on merit can coincide with those based on equality when everyone's contribution (input) to a common output is of equal value (either actually or supposedly), which \citet[115]{leventhal_distribution_1976} called the ``confluence of equity and equality norms''.
But then the \emph{principle} of distribution is still merit, not equality.
Investigating ``backward reasoning'' from visible rewards to unknown performance, \citet[181]{bierhoff_social_1986} reported that unequal rewards would activate the equity script, whereas equal rewards would activate both the equality script \emph{and} the equity script, not the equality script alone. 

There is a good reason to choose an equal distribution -- equality is not a fallback here. 

\subsubsection{The Communicability of the Chosen Distribution}
But it would be a mistake, however, to contrast these two allocation norms too sharply. 
\citet{leventhal_distribution_1976} discussed in detail the role of an allocator in groups or organizations.
Allocators would usually choose a compromise satisfying both norms partially (p.~114).
They must balance several dimensions, among them the productivity of the group, avoidance of conflict, and their own approval and acceptance by the group members (authority).

The chosen allocation rule is strongly contingent on communicability considerations:
when allocators can conceal their allocation decision to the group members (secrecy), they are inclined to follow the equity norm, for reasons of justice, productivity, and reduced group conflict due to a lack of social comparison \citep[111]{leventhal_distribution_1976}.
When group members are strongly interdependent, when they perceive themselves as ``similar'', when their success can only be credited to the whole group, and when group productivity can be fostered by keeping up a ``solidary spirit'' within the group, several authors reported a preference for an equal distribution (\citealt[108ff.,]{leventhal_distribution_1976}, \citealt[418ff.,]{greenberg_approaching_1982}, \citealt[147]{deutsch_distributive_1985}).

It seems that an individual's contribution to a genuine group achievement must stand out \emph{conspicuously} to justify and enforce an unequal distribution of rewards. 

\subsubsection{The Communicability of Costs and Benefits}
Social exchange includes tangible goods or services, but also less ``concrete'' resources like status or information \citep[Resource Theory,][]{foa_resourcetheory_2012}.
Less concrete does not mean less valuable or not targeted at material gain (which from the sociobiological point of view is the decisive outcome). 
\citet[ch.~9]{frank_luxury_2000} vividly described that status gain is not an objective in itself, but functional: A means to achieve higher income, better health, or more promising social relations (esp.\ in mating).

Long-term cooperation is contingent on many circumstances, so it can be advantageous to forego immediate benefits.
\citet{mikula_justice_1980} and \citet{schwinger_threeprinciples_1980} reported that in groups a ``politeness ritual'' would prevail when the group members themselves were to propose a distribution rule: 
High performers tended toward an equality rule and low performers toward a proportionality rule, both seemingly at odds with their material interests.
Yet \citet{bierhoff_social_1986} cited several studies where high performers tended toward more equal distributions if future interactions could be expected (but to equity otherwise).
This can be regarded as a self-binding (investment) since reciprocity by definition depends on repeated interaction. 
Social harmony, less conflict, and a high degree of group cohesion are less visible benefits, justifying very visible costs.

\subsubsection{The Limits of Performance-Related Pay}
\citet{deutsch_distributive_1985} markedly criticized a lack of empirical evidence that, within a group, \emph{individual} productivity would be increased by equitable rewards, making a case for equality as a default reward system in groups. 
Indeed, `Performance Related Pay' as the periodic adjustment of wages and salaries to individual performance is ever since plagued with countless difficulties, especially when job responsibilities are ambiguous (e.g., \citealp{isaac_performance_2001}, \citealp{perry_back_2009}, \citealp{breisig_entgelt_2003}).
As discussed in Section \ref{justsec_reciprocity_principle}, merit is uncontroversial only when it can be related to objective criteria. 

But Deutsch underestimated the significance of performance regarding \emph{routinely} paid wages and salaries.
His experiments lack an `employer' who has a personal interest in high performance at reasonable costs and who competes for the best and brightest with other employers. 
Unequal pay according to qualification and performance is general practice, and \citet{greenberg_approaching_1982},  \citet{leventhal_distribution_1976} as well as \citet{walster_equity_1978} provide many of the issues employers would consider when choosing levels of pay. 
Performance related pay in this sense is used to find (and keep!) the right people for the right job.
The interesting feature of a market price is that it is a collective, rather neutral assessment of performance and achievement, a condition usually not met in small groups.

\subsubsection{The Case of Israeli Kibbutzim}
Kibbutzim are voluntary rural communities in Israel that share an egalitarian ideal.
They are characterized by communal living with equal sharing of outcomes, no private property, and no cash exchange inside the kibbutz.
The main rationale for the existence of kibbutzim is economic risk-sharing, average group size is about 440 \citep[for this and the following cf.][]{abramitzky_lessons_2011}.

Kibbutzim largely face three threats, as predicted by economic theory and empirically verified: the exit of more productive members (brain-drain), the entry of less productive members (adverse selection), and shirking in the communal work (free-riding). 
In effect, kibbutzim must set straight the reciprocal calculation for each member, be it a high or low performer. 
This includes incentives and sanctions, some of them nonmonetary, feasible only in small groups: tight personal control, social prestige etc. 
Kibbutz members must be constantly reminded, by education and ideology, that they are investing into a common project.
While kibbutz members see themselves as counter examples to \emph{economic man}, Abramitzky showed that economic principles do not end at the kibbutz gate.

\subsection{Equality of Access}
\label{justsec_equality_access}

Application processes are generally characterized by an equality of access.
It would contradict the idea of an ``application'' when applicants with a \textit{prima facie} claim on the good being allocated are treated differently before the decision process \citep[cf.][ch.~5]{miller_principles_1999}. 
But the decision is basically about merit or guilt. 

Equality before the law primarily affects the ability to be indicted, the possibility to ``apply'' for a trial, no matter what status or ``deserts'' one has. 
During the trial, procedural justice requires all defendants to be treated impartially, but substantial justice requires the opposite.
The exact purpose of a trial is to determine objectively the extent of guilt and to find out what the just deserts are.

Likewise with political offices. 
The goal of every election of representatives is to bring only the best and brightest in offices, not anyone.
But we admit everyone to be elected, again for reasons of self-binding, and merit is assessed afterwards according to results.
If in doubt the person will be unelected on the next occasion.
But if there is a candidate of whom we know that she is the best, she will certainly win the election, yet not according to criteria determined in advance, but based on a collective assessment. 

The same holds true for all other applications, be it a new job or a place at university -- equality is equality of opportunity, the rest is about merit, but this will be assessed afterwards. 
It is a common misunderstanding made in philosophical debates on justice when asking ``Who deserves a job?'' \citep[cf.][ch.~8: ``Deserving Jobs'']{miller_principles_1999}: already the \emph{hiring} is regarded as a reward. 
But the rewards only come later, as monthly payment for achievements or as high grades for good exams, i.e., according to merit. 
The successful application as a chance of a true test can be viewed as a reward in a factual sense, but not in the systematic sense discussed here.

\subsection{Political Equality}
\label{justsec_equality_political}

Suffrage or membership in political bodies has become ever more inclusive because ever more reasons for inequality revealed as dysfunctional in an ever more liberal society had to be dropped. 
In fact, no reason has remained except a (contested) minimum age, and the self-binding of equal dignity and respect for all humans is now widely accepted as a minimal standard \citep[cf.][sec.~2.3]{gosepath_equality_2011}.
Most societies would profit if only ``thoughtful and well-educated'' people could vote (\emph{noocracy}), yet try to operationalize that.
Already the idea seems absurd, and all respective historical attempts have failed spectacularly.
Especially when placing people in positions where they are to expend other people's money (e.g., taxes for public expenditure) or decide over access to advantages we must be very cautious with conflicting interests.
It could put the fox in charge of the henhouse, though, when giving the wealthy a higher share of votes, and the same holds true for all other reasons once deemed good.
``One person, one vote'' therefore is not rooted in the idea of equality as such, but in an effective self-binding on a highly sensitive field.

\subsection{Summary}
Equality is not a principle of distributive justice but a fallback, chosen for a lack of good reasons to choose otherwise.
Like need, equality can be a case for entitlement.
Of course, it is contested what good reasons are (or are not), but for many important cases they are uncontroversial. 
Some reasons once good have lost their plausibility and have been removed from the ``societal pool'' of arguments. 
Other cases often stated as examples of ``genuine'' equality have revealed their reciprocal nature and their compatibility with utility maximization, though sometimes only as a cost avoidance strategy. 

Inequality is often identified with unfairness, but the opposite seems to be general practice: people prefer fairness to equality \citep{starmans_whypeople_2017}.

\section{Practices of Justice}	
\label{justsec_practice}

\subsection{Simplicity and Class Formation}
\label{justsec_practice_simplicity}

Gigerenzer’s \citeyearpar[199]{gigerenzer_gutfeelings_2007} statement is concise: ``Simplicity is the ink with which effective moral systems are written.''
He refers to the Ten Commandments as being incomplete, but clear, short and focusing on the essential.
Similarly, the rules of Bentham and Kant are clear and simple and have lasted for centuries. 
Some very interesting observations regarding simplicity can be found in Schelling’s classic ``The Strategy of Conflict'' \citeyearpar{schelling_strategy_1960}.
\emph{Focal points} are outcomes where mutual expectations easily converge because they ``enjoy prominence, uniqueness, simplicity, precedent [\ldots]'' (p.~70). 
Schelling points to the ``strong magnetism in mathematical simplicity'' in international bargains and emphasizes the ``remarkable frequency with which long negotiations [\dots] converge ultimately on something as crudely simple as equal shares, shares proportionate to some common magnitude [\ldots] or the shares agreed on in some previous but logically irrelevant negotiation'' (p.~67).
This makes clear that the \emph{communicability} of the outcome is decisive. 
A focal point is a good reason for both sides, as are simplicity, symmetry etc. 
If some better reason can be found, then this will be the defendable, ``reasonable'' basis of distribution.
In the debate about distributive justice, the principles of merit, need, and equality can be viewed as focal points, which might explain their prominent scholarly support in social psychology (cf.\ Section \ref{justsec_introduction}). 

Another important way toward simplicity is class formation.
Ideally, a just judgement would reflect the `true‘ relation of costs and benefits of the case in question, but often neither is clearly determinable. 
A (mathematical) class is a group of elements that are not equal but can be viewed as \emph{equivalent} regarding a certain aspect. 
Classes are abundant: pupils get grades, students make bachelor‘s and master’s degrees (and get grades, too), boxers box in weight classes, worker’s payment is according to wage grades, VAT has certain tax rates, letters‘ postage is according to weight classes and distance zones, parking meters charge full hours etc. 
The point here is that classes by definition are \emph{distinct} to make unequal things comparable, be it for transparent pricing or equal chances or due assessment.
Classes are used to make justice simpler and more efficient, but they have a few snares. 

Take for example how cars are taxed.
The bigger the car the higher the taxes, since bigger cars demand more public resources. 
But what does `size‘ mean? 
Different countries have different approaches (cf.\ Wikipedia: ``Road tax''), but they have in common that they define classes as distinct intervals of easily determinable values that can serve as proxies for size, e.g., engine displacement like 2001cc--2500cc, 2501cc--3000cc etc.\ as to have only a manageable number of classes and therefore a relatively simple system. 
Is it just?
Well, it is contingent, but at least it is \emph{not unjust}, practicable, communicable, and reliable. 

Other important classes are nations (one of the most delicate class formations at all), for making very generalized reciprocity manageable, or the fiscal year, for finding a compromise between long-term fluctuations of profits (or income) and the need to come to terms with the past. 
The fiscal year is a \emph{practicable} compromise, not the only possible, but surely a focal point, in contrast to nations where focal points (ethnicities) are often ignored and continue to be sources of endless struggle. 

One of the most obvious problems of classes besides their general contingency are `absurdities‘ along the class boundaries.
It is easy to determine the engine displacement of a car, but difficult to reason whether a good is a `necessity‘ and therefore to enjoy a lower VAT rate than a `luxury good‘.
Obvious cases are easy, like bread and jewelry, but what about sweets and clothes?
Is it helpful to further differentiate between `necessary‘ and `superfluous‘ sweets and clothes?
Tax regulations and their commentaries are infamous for their length and complexity, and some of them \emph{are} absurd. 
But most of it is unavoidable, since generally the executive is bound to a strict procedural justice with transparent rules agreed upon in advance. % (cf.\ Section \ref{justsec_implementation_procedural}).
These absurdities do not challenge the concept of class formation in general. 

Justice is arduous, a never-ending balance between (brute) simplicity and (endless) differentiation. 

\subsection{Ideal Justice and Other Criteria}
\label{justsec_practice_othercrit}

There are some places in Miller’s book \citeyearpar{miller_principles_1999} where he casually, but confidently distinguishes justice and other criteria:
``Even when genuine moral considerations -- as opposed to simplicity or convenience -- guide the choice of an allocation or a procedure, it may not be obvious whether \emph{justice}, specifically, is what lies behind the respondent’s endorsements’’ (p.~47, original emphasis).
Some lines further down he contrasts justice and social utility, somewhere else justice and efficiency (p.~216) or justice and consent (p.~104f.). 
These remarks are quite spectacular. 
He never defines what `justice’ is, and nobody else does so (at least not successfully) beyond ``giving everyone his or her due’’. 
Miller has an intuition of an ideal justice, and this intuition is generally reflected in the debate in terms like ``genuinely moral’’, ``legitimate’’, ``adequate’’, ``justified’’, or simply ``fair''.
\citet[p.~52]{leventhal_whatshould_1980} illustratively contrasted ``fair'' and ``quasi-fair'' behavior, while \citet{greenberg_whyjustice_1982} would term it normative and instrumental justice. 
A famous example for this justice intuition is \textit{Radbruch's formula}, developed after World War II in reaction to German National-Socialist law practice \citep{radbruch_unrecht_1946}.
Radbruch objected the positivist identification of law and justice and emphasized the role of law `to serve justice'. 
But besides obvious violations -- is pitting ideal justice against other criteria helpful in practice?

Viewed from the evolutionary perspective, long-term utility (functionality) is the \emph{decisive} dimension of justice, but humans are not perfect.
Yet \emph{because} they are maximizers of utility, ostensibly utilitarian (functionalist, short-term) considerations can go awry. 
But the goal is the long-term participation in the social game.
On the one hand, we must check ourselves via individual self-bindings: good intents, conscience, building up reputation, contracts, reliability etc. 
On the other hand, we must efficiently solve the free-riding problems of 1st and 2nd order: to recognize and sanction norm violators \citep[p.~228]{voland_soziobiologie:_2013}.
Practical justice \emph{must} be efficient.
It would not help to use up the resources that are to be defended by justice. 
Yet we must be careful to not to overshoot the mark while recognizing and sanctioning norm violators: ``Shoot first, ask questions later’’ is a safe road to an unjust society. 
Therefore, a large part of our numerous self-bindings is not individual, but collective (institutional): fair trials, transparent administration, democratic decisions -- procedures to avoid premature decisions biased to our personal advantage at the expense of others and therefore maleficent in the long run. 

To have an ideal of justice is necessary for analyzing `true’ merit or guilt and for demarcating it from additional criteria and functionalist intentions that try to push efficiency or outcomes (very popular: `jobs’) into the focus of the debate while neglecting the basic problems. 
But in practice a principled distinction between `noble’ justice and `mundane’ considerations is difficult. 
Ideal justice is the default here, but good reasons can override the default.

\section{The Implementation Problem}
\label{justsec_implementation}

\subsection{From Moral Will to Institutions}
\label{justsec_implementation_institutions}
How is a Theory of Justice successfully implemented? 
Traditional moral philosophy and common sense focus on individual actions, assuming a duality of ``genuinely moral motives'' and ``all other motives'' (cf.\ \citealt{homann_sollen_2014}, critically discussing this discourse; cf.\ also \citealt[51]{trivers_altruism_1971}). 
Moral weakness is a weakness of moral \emph{will}, the answer being ``moral rearmament'', i.e., developing better moral reasoning to resist temptations.

\citet[451]{ulrich_integrative_2016}, for example, \emph{contrasts} economy and morality, and morality would require a breach of economic logic. 
\citet{homann_2003_anreize, homann_sollen_2014} is a critic of such a dualistic opposition: 
Moral behavior must be advantageous for the individual if he or she is not to be exploited by others who are less moral and would initiate a moral downward spiral. 
Any successful implementation of ethical rules therefore must be compatible with economic incentives. 
\citet[226]{homann_2003_anreize} warned against regarding implementation ``only'' as a practical problem: implementability precedes moral validity. 
An ethic that presents unrealistic challenges to its participants is not impractical but \emph{unethical}.

We can add another dimension: 

\textbf{Enforceability}: helpless $\sim$ liberal $\sim$ coercive

Justice must be reluctant, because only in liberty can people develop to the full, deciding themselves what is best for them.
Self-appointed guardians of virtue may not restrict the freedom of others, and paternalism or ``nudging'' are rightly criticized, as is governmental overregulation. 
But justice cannot be tame. 
The collective has a right to define justice and to enforce it. 
Liberty without coercion (anarchistic self-control) is unrealistic because people can drift far into self-righteousness \citep{haidt_righteous_2013} and make their own rules -- especially when rich or ``big''.

\citet{brennan_reason_1985} made a distinction between \emph{choices within rules} (choosing actions) and \emph{choices of rules} (designing constitution and laws).
\citet[62]{homann_sollen_2014} termed these \emph{moves in the game} and \emph{rules of the game}.
The rules of the game are the institutional framework governing the economic behavior in society -- in short: the ``economic order'', a term coined by ordo-liberalism, a neo-liberal movement in Germany in the mid-20th century \citep{eucken_foundations_1992}.
\citet[9]{homann_2003_anreize} locates the \emph{systematic} position of ethics in modern (anonymous, pluralistic) societies in this economic order.
The \emph{rules} must be moral as to enable the ``players'' to be guided in their moves only by economic considerations, since economic profits are a valuable indicator of sources of societal wealth \citep[99]{homann_sollen_2014}. 
It is only this compatibility of sanctioned rules and economic incentives that makes the system inherently stable.
The most important rule is to guarantee fair competition in market economies \citep{eucken_foundations_1992, homann_2003_anreize, mankiw_economics_2011}.
Individual ethics are nevertheless still important for filling the many gaps left out by the economic order.

\subsection{The Dark Side of Competition}
\label{justsec_implementation_competition}
The significance of fair competition for trade is widely acknowledged, yet current political practice often looks different. 
Governmental overregulation is only one part of ``economic policy'': attempts to influence or correct the outcomes of ``free'' markets, with subsidies, protectionism, welfare programs, growth policies, comprehensive labor and employment law etc.
So, what is going wrong with fair competition?

Worldwide, economic growth is a primary goal of politics. 
\citet{richters_contested_2019} recently argued that a political growth imperative exists \emph{because} politics would strive for justice, and that much of what is considered to be fair competition is quite the opposite. 
They point to the fact that productivity increases due to technological progress are not only the result of ``good ideas'' but have an underestimated material basis \citep{ayres_economic_2009, kummel_second_2011, ayres_underestimated_2013}. 
Accordingly, they distinguish a ``genuine'' form of competition based on performance, when extraordinary personal abilities and efforts (``merits'') are honored, and an innovation competition, when not the innovation as such but mainly its resource use leads to a better cost-benefit ratio of products and services and, accordingly, to market success \citep[134]{richters_contested_2019}.
Technology then undermines the meritocratic principle by literally using resources not based on merit.
This systematic distortion of fair competition allows technical occupations not only to bias the income distribution toward their own benefit but is responsible for the continuous threat of ``technological unemployment''. 

A political growth imperative in their view is created by the interplay of three factors (p.~133): (1)~increases in labor productivity, leading to technological unemployment, (2)~the societal obligation to guarantee at least a minimal standard of living for everyone (``need''), and (3)~the meritocratic principle as a fundamental social norm that sets limits to redistribution between those earning a sufficient income and those who are not.

\subsection{Concentrating on ``Non-Merit''}
\label{justsec_implementation_nonmerit}
Section \ref{justsec_reciprocity_principle} described the wrangling over the meaning of merit. 
Participants in a study by \citet[45]{neckel_umkampfte_2008} found it easier to agree on what merit is \emph{not}.
This suggests using the \emph{exclusion} of \emph{non-merit} as an effective implementation of the meritocratic principle.
From an Aristotelian perspective this would mean not struggling with the broad mean between extremes, but concentrating on the extremes themselves.

\citet{richters_making_2021} presented a possible solution to the problem of how to implement the meritocratic principle regarding the side conditions of effectiveness, efficiency, communicability, and liberty. 
In economics, undeserved incomes without a corresponding creation of wealth are called ``economic rents''.
Thus, justice in modern, democratic market societies would first and foremost mean to institutionally drain the wellsprings of such economic rents. 
They extensively discuss what in their view are two important sources of economic rents: (1) income shares which are substantially based on resource-intensive technologies, and (2) land rents, where a value generated by society is sold on private account. 
Appropriate institutions are caps on the extraction or import of non-renewable natural resources and a land value tax.
Without economic rents, market economies could come closer to the goal of just self-regulation \citep{richters_marktwirtschaft_2019}.

\section{Discussion}
\label{justsec_discussion}

\subsection{Reconciling Altruism with Reciprocity}
I have argued that the sociobiological notion of reciprocity can be regarded as the ideal of justice, and that an overwhelming, though often inconspicuous evidence indicates that reciprocity indeed governs all social relations -- with the caveat that the \emph{term} reciprocity seems to be an inappropriate framing for a substantial number of social situations (cf.\ Figure \ref{justfig_reciprocity}). 

The more generalized and the less obvious reciprocity is, the more difficult it is to argue that certain actions are still part of an individual maximization of utility in the long run.
But we must take seriously the sociobiological account that altruism in its selfless interpretation could not have been selected for in the evolutionary process. 
Although the term altruism may still have some justification in everyday language, for systematic reasons we should interpret altruistic acts as ``blind investments'' into society with a questionable ``profitability'' -- hence, assuming reciprocity. 
With reference to \citet{sahlins_sociology_1965} I have suggested to view such acts as ``very generalized reciprocity''.
We do not ``need'' altruism to explain stable patterns of just cooperation in societies. 

\begin{figure}[t]
\hfil\includegraphics[width=\columnwidth]{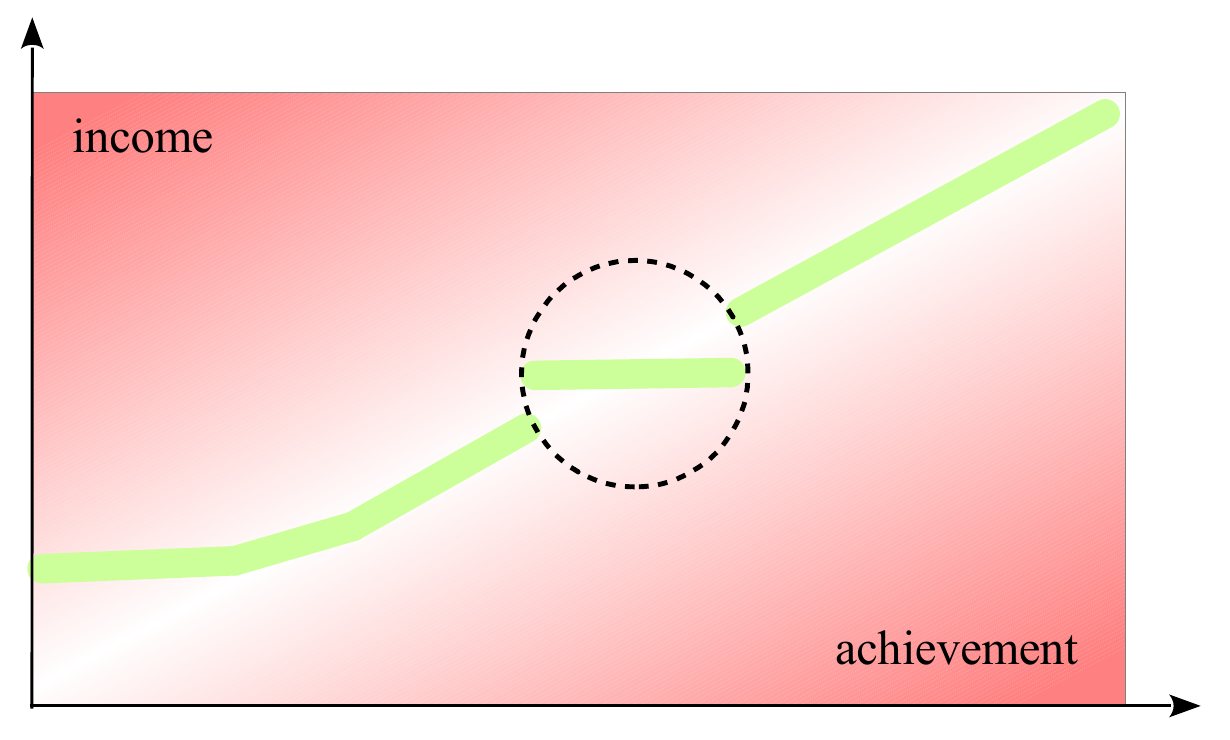}\hfil
\caption{\label{justfig_realjustice}The Practice of Social Justice}
\end{figure}

In practice, justice is a social bargain and an optimization problem, with some of the effects depicted in Figure \ref{justfig_realjustice}: 
Achievements and income serve as proxies for deserts and rewards, but our fellow citizens should not fall below the socio-cultural breadline. When individual efforts in groups are roughly comparable, their members may not argue about subtleties but prefer equality.

If we assume that the central goal is utility maximization, then balancing reciprocal expectations becomes only one goal among others that affect costs and benefits. 
Often maximizing net utility means minimizing (social, invisible) costs rather than maximizing (material, visible) benefits.

\subsection{The Problems of Equity Theory}
To approach a general theory of human interaction was indeed the claim of Equity Theory, but the way the theory was once presented (and defended) was probably infelicitous.
Equity Theory would allow both signs (positive and negative) for inputs and outcomes, which made the concept confusing, and justified criticism was levelled where the meaning of inputs and outcomes had been repeatedly stretched only to achieve a balance, making the terms meaningless (\citealt[e.g.,][98f., 105f.,]{schwinger_threeprinciples_1980}, \citealt[p.~30,]{deutsch_distributive_1985}, \citealt[189]{reis_levels_1986}).
The mathematical formula presented conveyed the idea of precision and applicability but could never deliver on its promise \citep[cf.][]{folger_rethinking_1986}.

The basic problem I see is that the social notion of ``inputs'' and ``outcomes'' refers implicitly to the visible (communicable) preconditions and effects of social exchange.
The notion of costs and benefits with their definite sign and ``holistic'' meaning is much clearer.
There will never be a ``quantification'' of equity (just another word for full communicability), as demanded by \citet{adams_equity_1976}, when the terms of the underlying formula are difficult to quantify, strategically used, and stretched or ``tugged'' between conflicting interests.
But there could be a theoretical agreement on its underlying principle. 

\subsection{Monism as Preferable Approach}
Concerning the criticized ``unidimensionality'' of Equity Theory, I would regard it a virtue, not a failure, always keeping in mind that an operationalization of the meritocratic principle will only be possible by concentrating on non-merit.
We need unidimensional theories to discover the relevant forces, i.e., a (strong) default and (weaker) deviations. 
That does not mean that social action gets uniform or ``deterministic'', or that society would be ``simple and free of contradictions'' (\citealp[p.~105,]{schwinger_threeprinciples_1980}, ironically). 
Rather we have seen that highly sophisticated strategies and arguments are used, often exactly for diverting the focus from the main dimension reciprocity to important other criteria. 
But basically, these remain to be ``other'' criteria. 

\subsection{The Relation of Distributive and Procedural Justice}
t would be interesting to revisit the relation of distributive and procedural justice with the idea of reciprocity as the primary principle of justice.
When outcomes are decisive, what is the systematic role of procedures?
Social psychology regards both to be important dimensions without hierarchizing them \citep{lind_study_2020}.
I would rather assume that insisting on certain procedures could be better explained as self-bindings for maintaining impartiality, avoiding biases on the part of the allocator.
This would entail a primacy of outcomes, making procedures ``only'' part of the communication about justice and their outcomes. 

\subsection{The Social Construction of Merit}
Reciprocity -- especially regarding income for work -- translates into the social norm \emph{Meritocratic Principle} whose basic idea is to ensure rewards for your own achievement.
Strictly speaking, there is nothing like ``your own achievement''.
The principle of the natural cycle is that all organisms impropriate calories ``at the expense of others'', usually entailing the death of the organisms that have built them up. 
All creatures must choose their food (prey) according to whether its acquisition delivers more calories than it costs. 
To paraphrase an adage of Marxism, we all ``make more calories from calories'', and the ``profit'' (excess calories) is used up for self-preservation and also (sociobiologically: first and foremost) for reproduction. 

Therefore, Rawls and many others are not plainly wrong in their assumption that ``merit'' is a social construction.
But it is not an arbitrary one. 
First, merit has a non-negotiable core, especially when time or quantities are the objective yardstick of achievement or when some people achieve results that others cannot, no matter how much effort they invest. 
Second, this construction has a deeper social meaning.
Merit is the normative basis of the division of labor since nobody can achieve unboundedly (this is why we divide labor). 
Effort (i.e., hours spent on work) is an income-limiting factor, because only in cooperation with others can the talents of the few unfold and create specialists who deserve higher wealth.
Their productivity is contingent on their social environment (team, firm, society, \ldots) -- hence, how is ``high performance'' to be attributed?

The meritocratic principle has a strong egalitarian element that is often not allowed for in the debate.
There has been considerable resistance to the notion that the relationship between income and economic deserts in unregulated markets is just \citep[e.g., the discussion in][]{olsaretti_liberty_2004}, and I fully agree. 
But \citet{richters_making_2021} have exhibited a way out of these troubles by showing how justice in markets can be achieved with few but strong and consistent measures.
One of these are limits to inequality because inequality can develop a runaway dynamic far beyond any merit.
From the collective perspective, endless accumulation is (and always has been) a social misunderstanding (though not from the individual perspective of course).

\subsection{Focusing on the Essentials}
A monist approach to justice based on biological facts and a common, basically uncontroversial practice has yet another advantage: it is a Theory of Justice that itself is normatively parsimonious. 
According to \citet{gigerenzer_gutfeelings_2007}, effective moral systems should be clear, short and focus on the essentials, and this holds also for the institutions of modernity. 
The current philosophical debate often looks different \citep[cf.][]{olsaretti_handbook_2018}: extreme positions try to avoid \emph{any} bias (e.g., political correctness) and compensate for \emph{any} perceived injustice (e.g., when couples prove unfertile and cannot have the children they wish).
Instead, the societal quest for justice should concentrate on the blatant cases, i.e., ``non-merit'', and should be content with results that are acceptable, i.e., ``not unjust''. 

\clearpage

\renewcommand{\bibfont}{\normalfont\footnotesize}
\addcontentsline{toc}{section}{References} 
\newrefcontext[sorting=nyt]
\printbibliography

\end{document}